\documentclass[pra,twocolumn,superscriptaddress,notitlepage,amsmath,amssymb,showpacs,floatfix]{revtex4-1}
\usepackage{graphicx,subfigure,amssymb,amsmath,amsfonts}
\usepackage{verbatim}
\usepackage{ulem}
\usepackage{multirow}
\usepackage{hyperref}
\usepackage[T1]{fontenc}
\usepackage{chngcntr}
\usepackage{bookmark}

\usepackage{dsfont}
\usepackage[usenames]{color}

\hypersetup{
    colorlinks=true,       		
    linkcolor=black,          
    citecolor=black,          
    filecolor=black,      		
    urlcolor=black,           
    runcolor=cyan
}

\DeclareSymbolFont{AMSb}{U}{msb}{m}{n}
\DeclareSymbolFontAlphabet{\mathbb}{AMSb}

\begin{document}
\title{Monitoring currents in cold-atom circuits}

\author{S. Safaei} 
\affiliation{Centre for Quantum Technologies, National University of Singapore, 3 Science Drive 2, Singapore 117543}
\affiliation{Majulab, CNRS-UCA-SU-NUS-NTU International Joint Research Unit, Singapore}

\author{L.-C. Kwek} 
\affiliation{Centre for Quantum Technologies, National University of Singapore, 3 Science Drive 2, Singapore 117543}
\affiliation{National Institute of Education, Nanyang Technological University, 1 Nanyang Walk, Singapore 637616}
\affiliation{Institute of Advanced Studies, Nanyang Technological University, 60 Nanyang View, Singapore 639673}
\affiliation{Majulab, CNRS-UCA-SU-NUS-NTU International Joint Research Unit, Singapore}

\author {R. Dumke} 
\affiliation{Centre for Quantum Technologies, National University of Singapore, 3 Science Drive 2, Singapore 117543}
\affiliation{Division of Physics and Applied Physics, School of Physical and Mathematical Sciences, Nanyang Technological University, 21 Nanyang Link, Singapore 637371}
\affiliation{Majulab, CNRS-UCA-SU-NUS-NTU International Joint Research Unit, Singapore}

\author{L. Amico}
\affiliation{Centre for Quantum Technologies, National University of Singapore, 3 Science Drive 2, Singapore 117543}
\affiliation{Majulab, CNRS-UCA-SU-NUS-NTU International Joint Research Unit, Singapore}
\affiliation{Dipartimento di Fisica e Astronomia, Via S. Sofia 64, 95127 Catania, Italy}
\affiliation{CNR-MATIS-IMM \& INFN-Sezione di Catania, Via S. Sofia 64, 95127 Catania, Italy}
\affiliation{LANEF {\it 'Chaire d'excellence'}, Universit\`e Grenoble-Alpes \& CNRS, F-38000 Grenoble, France}

\begin{abstract}
Complex circuits of cold atoms can be exploited to devise new protocols for the diagnostics of cold-atoms systems. 
Specifically, we study the quench dynamics of a condensate confined in a ring-shaped potential coupled with a 
rectilinear guide of finite size. We find that the dynamics of the atoms inside the guide is distinctive of the 
states with different winding numbers in the ring condensate. We also observe that the depletion of the density, 
localized around the tunneling region of the ring condensate, can decay in a pair of excitations experiencing a 
Sagnac effect. In our approach, the current states of the condensate in the ring can be read out by inspection 
of the rectilinear guide only, leaving the ring condensate minimally affected by the measurement. We believe that 
our results set the basis for definition of new quantum rotation sensors. At the same time, our scheme can be 
employed to explore fundamental questions involving dynamics of bosonic condensates.
\end{abstract}


\maketitle

\section{Introduction}
%
Nowadays, cold-atoms systems provide a tunable and flexible platform for studying quantum liquid behavior~\cite{Leggett}. 
With the advances in quantum technology, remarkable progress has been achieved in the field. Concomitantly, cold-atoms 
systems have provided new tools, devices and perspectives to explore other branches of physics. And lying deep in this 
framework is a new field of atomtronics~\cite{seaman2007atomtronics,amico2005quantum,Amico_NJP}. This field seeks to 
realize atomic circuits where ultracold atoms are manipulated in a versatile laser-generated or magnetic guides. An 
important goal of the field is to enlarge the scope of the cold-atoms quantum simulators to study fundamental aspects of 
quantum coherent systems. At the same time, atomtronics aims at fabrication of new quantum devices and sensors with enhanced 
control and flexibility, by exploiting special features of the neutral cold-atoms quantum fluid~\cite{barrett2014sagnac,
arnold2006large,navez2016matter,dumke2016roadmap}.

There has been much interest in the simple  circuit made of a bosonic condensate flowing in ring-shaped 
guides and pierced by an effective magnetic field~\cite{dalibard2011colloquium,wright2013driving,Ramanathan2011,Ryu2013,
eckel2014hysteresis,yakimenko2015,eckel2014hysteresis,hallwood2006macroscopic,solenov2010metastable,amico2014superfluid,
aghamalyan2015coherent,aghamalyan2016atomtronic,aghamalyan2013effective,Mathey_Mathey2016,haug2018readout}. 
We note, however, that the recent progress in the field allows us to access richer scenarios. Indeed, condensates 
can be loaded in basically arbitrary potentials with micron-scale resolution~\cite{Zupancic:16,haase2017versatile}. 
In addition, such potentials can be changed in shape and intensity at time scales of tens to hundreds microseconds, and 
therefore opening the way to modify the features of the circuit in the course of the same experiment (typically involving 
tens of milliseconds)~\cite{Gauthier:16,Liang:09,muldoon2012control,Boshier_painting}. Remarkable advances on the 
flexibility and control of cold-atoms quantum technology, in turn, has opened up exciting possibilities for atomtronics. 
First, micro-fabricated integrated circuits of cold atoms can be feasibly realized. Second, the very shape and functionality 
of the circuit can be changed dynamically during its operation in a virtually continuous way.

Here, we study an integrated atomtronic circuit to realize new protocols for the manipulation of quantum 
fluids in complex networks of cold atoms. Schematically, the circuit is assumed to be divided into two distinct 
but coupled parts: 'primary' and 'secondary'. We assume that the quantum fluid operates in the primary part of 
the circuit. Then we ask: Is it possible to gain information on the primary part by manipulating solely the 
secondary circuit? To answer this question, we study the dynamics of a simple setting: A bosonic condensate 
flowing in a ring-shaped guide tunnel-coupled to a rectilinear quantum well. In our circuit, the primary part 
is the ring-shaped condensate; the secondary part is the rectilinear guide. We see that the different current 
states in the ring correspond to distinctive dynamics of the condensate in the guide. Such a protocol could 
then be used to read out the current states in a quasi-continuous way, being limited mainly by the quality of 
the achieved BEC that operates in the primary circuit. 

\section{The Circuit Structure}
\begin{figure}[t!]
\includegraphics[width=0.35\textwidth]{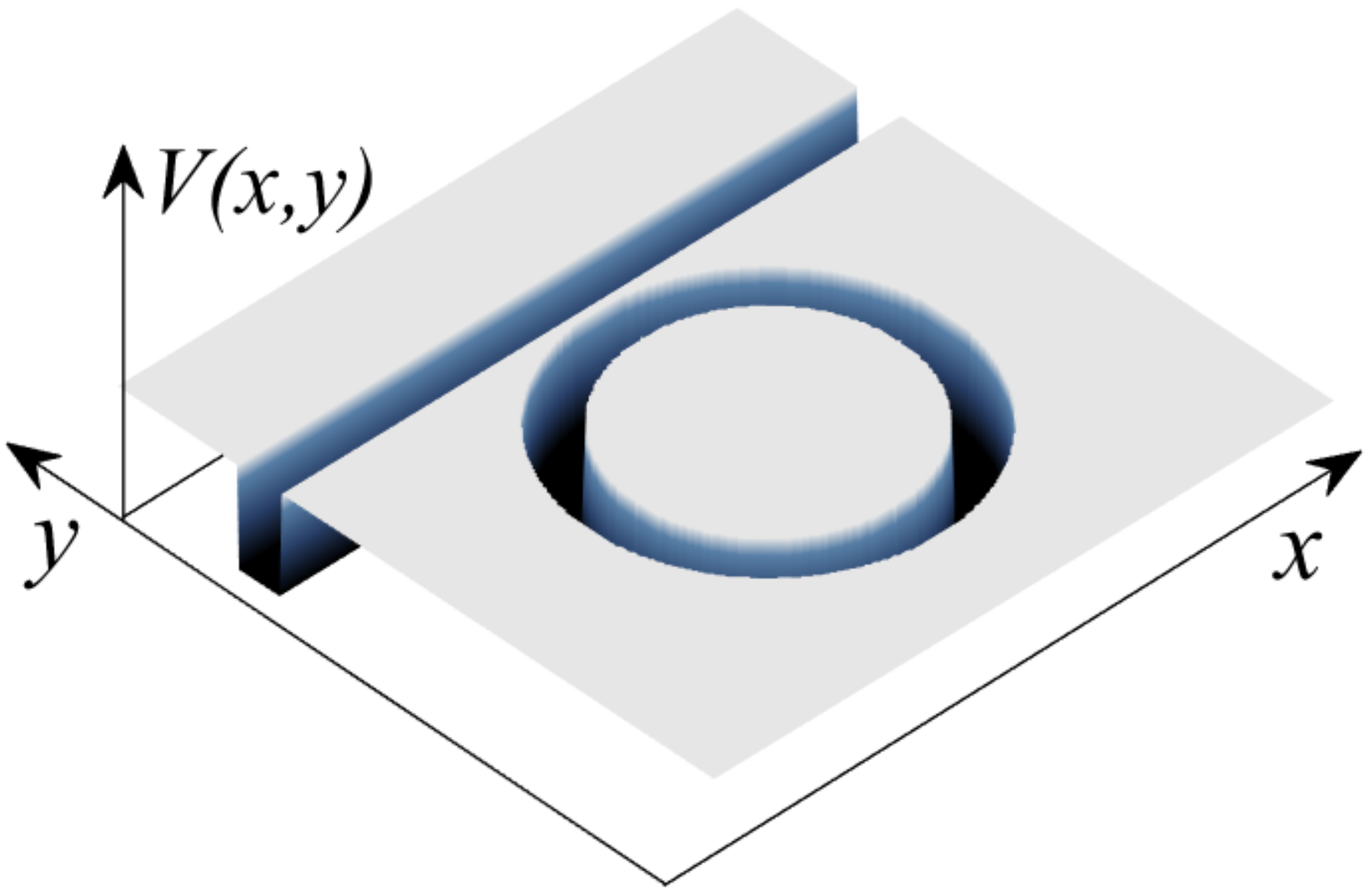}\\
\includegraphics[width=0.30\textwidth]{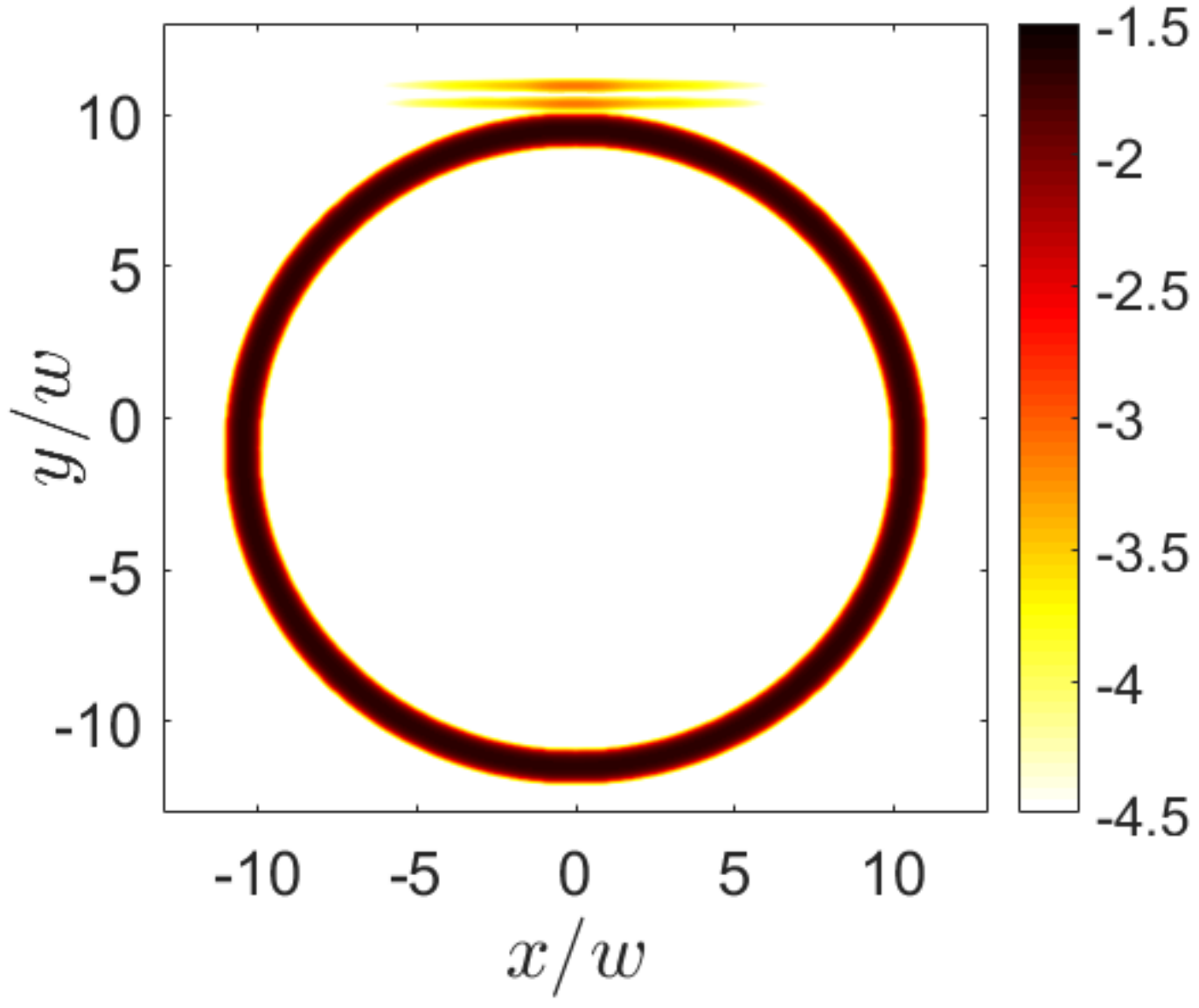}
\caption{Top: schematic drawing of the circuit of cold atoms consisting of a ring-shaped trap and 
a rectilinear waveguide in $x$ direction. The atoms are absorbed at the two ends of the waveguide. 
Bottom: the logarithm of the atomic density in the system when some atoms have tunneled from ring 
to waveguide. We consider a ring potential with width $w$ and radius $R=10.5w$ in a computational 
space spanned over $-20w\leq x \leq 20w$ and $-15w\leq y \leq 15w$. In $x$ direction, a layer with 
the width $\Delta=5w$ is dedicated to ABC in both sides and therefore, the physical space is limited 
to $-15w<x<15w$. The ring is centered at $(x_r,y_r)=(0,-w)$ and the waveguide, with the same width 
as the ring, is located at $y_g=10.7w$. For such a ring, artificial magnetic field strengths of 
$B\pi^2=0,\pm0.02,\pm0.04,\pm0.06$ result in winding numbers $\Omega=0,\pm1,\pm2,\pm3$, respectively. 
We also consider an initial total particle number of $N=6\times10^5$. Other parameters vary from 
case to case and their values are given when required. In the example presented here, the ring 
potential and the waveguide have same width $w$ and depth $U_0=20$. Atoms are stationary inside 
the ring ($\Omega=0$) and the atom-atom interaction strength is $u=2\times10^{-4}$. 
\label{fig:system}}
\end{figure}

%
The circuit is made of a two-dimensional ring-shaped condensate coupled to a two-dimensional rectilinear quantum well of finite length.
To paint a well-resolved circuit, we consider sharp potentials defined by step functions. The ring potential has radius $R$ and width 
$w$ centered at point $(x_r,y_r)$ and is defined with function $V_r(x,y)=-U_0$ when $R-\frac{w}{2}<r<R+\frac{w}{2}$ and is zero elsewhere. 
Here $r=\sqrt{(x-x_r)^2+(y-y_r)^2}$ and $U_0$ is the depth of the potential. A nearly resonant tunneling between the ring and waveguide 
is achieved for waveguide and ring with  the same width $w$ and depth $U_0$. The waveguide potential, placed at distance $y_g$ from the 
$x$ axis, is defined as $V_g(x,y)=-U_0$ when $y_g-\frac{w}{2}<y<y_g+\frac{w}{2}$ and is zero elsewhere (Fig.~\ref{fig:system}).

We assume that the dynamics of the BEC is governed by Gross-Pitaevskii equation (GPE) and we write, {in terms of} dimensionless {quantities},
\begin{eqnarray}
\label{eq:GP}
\begin{aligned}
& i\partial_{\tilde{t}}\tilde{\psi}(\tilde{\vec{r}},\tilde{t})=\\
& \left(
(-i\tilde{\vec{\nabla}}-\tilde{\vec{A}}(\tilde{\vec{r}}))^2+
\tilde{V}(\tilde{\vec{r}})+
N\tilde{u}|\tilde{\psi}(\tilde{\vec{r}},\tilde{t})|^2
\right)\tilde{\psi}(\tilde{\vec{r}},\tilde{t}),
\end{aligned}
\end{eqnarray}
where the dimensionless quantities are: 
$\tilde{\vec{r}}=k\vec{r}$, $\tilde{t}=\omega t$, $\tilde{\vec{\nabla}}=(\partial_x\hat{x}+\partial_y\hat{y})/k$, 
$\tilde{V}=V/E$, $\tilde{u}=2mu/\hbar^2$ and $\tilde{\psi}=\psi/k$ ($m$ is the mass of particles and $\hbar$ is 
the reduced Planck constant). The  recoil energy $E=\hbar^2 k^2/(2m)$, $k^{-1}=w/\pi$ and $\omega^{-1}=\hbar/E$ 
serve as the units of the energy, length and time, respectively. With our choice of the scaling units $\tilde{w}=kw=\pi$. 
The parameter $u=\frac{4\pi\hbar^2}{m\delta}a_s$ is the strength of the interaction in a two-dimensional system with 
s-wave scattering length $a_s$ and $3$D-to-$2$D scaling factor $\delta$.
 
The two-dimensional vector $\tilde{\vec{A}}=\vec{A}/(\hbar k)$, with $\vec{A}(\vec{r})=\frac{B}{2}(-(y-y_r)\hat{x}+(x-x_r)\hat{y})$, 
is the artificial gauge field resulting in an effective magnetic field with strength $\tilde{B}=B/(\hbar k^2)$ in $\hat{z}$ 
direction, and flux $\Phi=B\pi r^2$. With $\Phi_0=h$ being the the flux quantum, the winding number for the atoms at radius $r$ from 
the center of the ring reads as $\Omega=\text{int}(\Phi/\Phi_0)=\text{int}(\tilde{B}\tilde{r}^2/2)$.
Finally, we consider normalized (scaled and non-scaled) wavefunction, $\int d\vec{r}~|\psi|^2=1$, in the computational 
space and a total number of particles $N$.
Hereafter, we will work with dimensionless quantities and scaled GPE (\ref{eq:GP}) while dropping the tilde from the 
notation for convenience.

The atoms, which tunnel from the ring into the waveguide, spread in all directions and could reflect 
from a physical or computational boundary. Here we are interested in the case where atoms flow freely 
in the $x$ direction inside the waveguide. This scenario represents a physical system in which the atoms 
are absorbed, by detectors for instance, placed at the two ends of the waveguide or one in which the 
waveguide is sufficiently long so that there is no reflection in $x$ direction for the duration of observation. 
For this purpose we will apply absorbing boundary condition (ABC) in $x$ direction, minimizing  the atoms' 
reflection from the endpoints of the guide.

Indeed, there are different methods to apply ABC. Here we use a common method that makes use of an extra 
damping potential applied in a layer from the boundaries~\cite{jungel2010time,antoine2013computational}: 
the absorbing potential is equal to zero in the physical region where $x_{L}<x<x_{R}$ and is defined as 
$V_{ABC}=\frac{-iV_0}{\Delta^2}(x-x_{L/R})^2$ when $x_{L}-\Delta \leq x \leq x_{L}$ or $x_{R} \leq x \leq x_{R}+\Delta$. 
Here, $V_0$ is the strength of the absorbing potential and $\Delta$ is the width of the solely-computational 
region in which ABC is applied. We do not apply any ABC in $y$ direction. We note that the ABC is applied 
only during the real time evolution while for imaginary time evolution (used to compute the ground state 
of the system) the layers beyond $x_R$ and $x_L$ are treated as the usual computational and physical space. 
The waveguide potential $V_g(x,y)$ is also defined for $x_{L}-\Delta \leq x \leq x_{R}+\Delta$.

\section{Results}
%
We assume that the BEC is initially in the ground state corresponding to a circulating state of the atoms in the 
ring-shaped potential. Then, the gauge potential is switched-off and the trapping potential is quenched in such a 
way that the initially empty waveguide is turned on, next to the ring-shaped condensate. The atoms then tunnel from 
the ring into the waveguide.

Starting from the ground state of the atoms \cite{SupMat} inside the ring potential with the depth $U_0=20$ 
and atom-atom interaction strength $u=2\times10^{-4}$, for $\Omega=0,\pm1,\pm2,\pm3$ and an ABC with $V_0=20$, we let the 
atoms tunnel from the ring into the waveguide which has the same depth and width. We specifically monitor three quantities 
inside the waveguide in time: the total number of atoms $N_{tot}=N\int dy \int dx |\psi(x,y,t)|^2$, the net flux of particles 
in $x$ direction $\int\int J_x=\int dy \int dx J_x(x,y,t)$ where $J_x=-iN(\psi^*\partial_x\psi-\psi\partial_x\psi^*)$ 
is the $x$ component of the atomic current, and finally, the position of center of mass in $x$ direction 
$\langle x\rangle=\int dy \int dx \psi^*(x,y,t)x\psi(x,y,t)$. All integrals are taken over the waveguide area.

Notice that, largely due to atom-atom interaction inside the ring, the geometric resonance between ring and guide may be 
lifted. Accordingly, we find that density profile of the atoms in $y$ direction of the waveguide clearly displays that 
first excited state in the waveguide with energy $E_2$ (bottom panel in Fig.~\ref{fig:system}) is occupied. 

\begin{figure}[t!]
\includegraphics[width=0.49\textwidth]{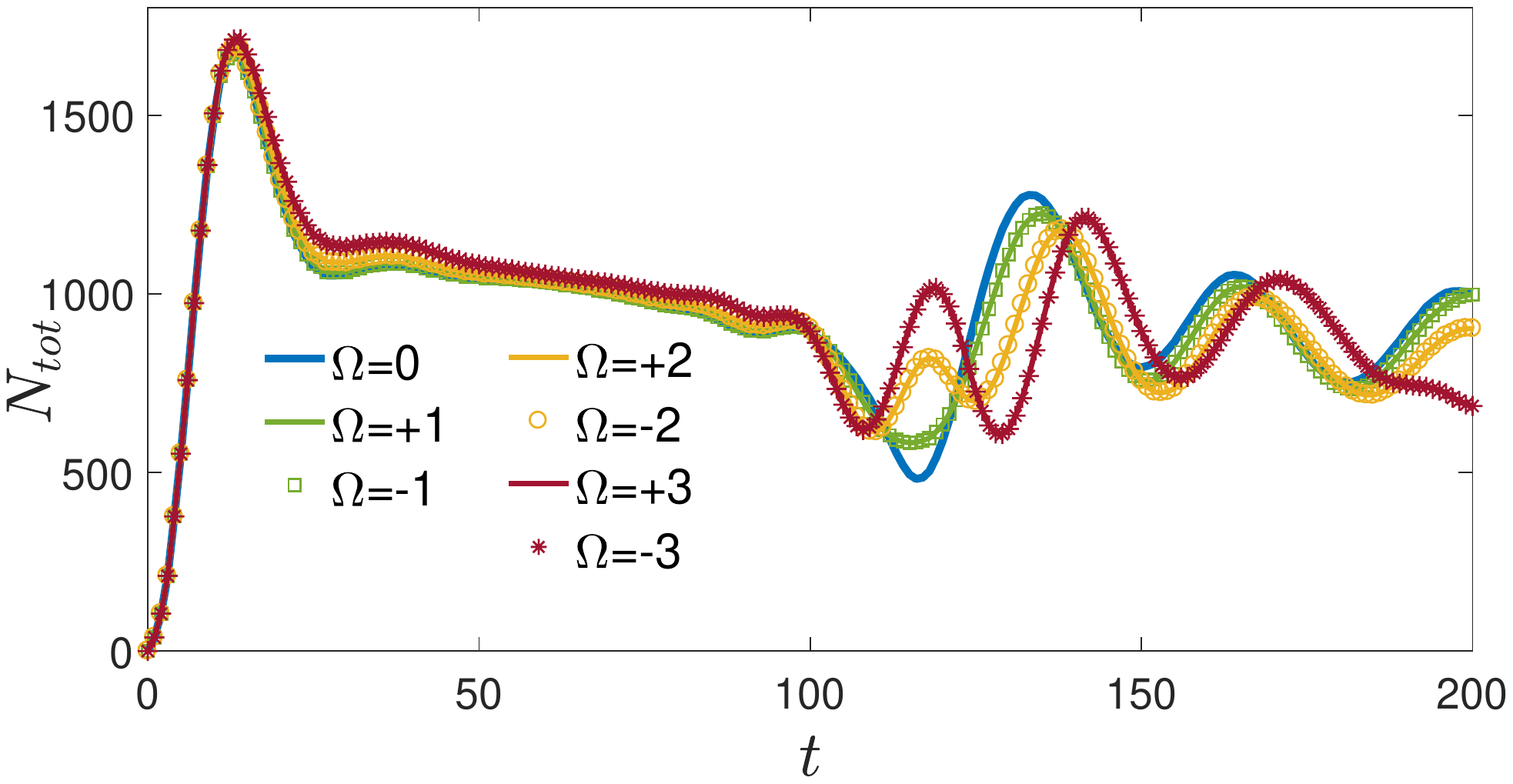}\\
\includegraphics[width=0.49\textwidth]{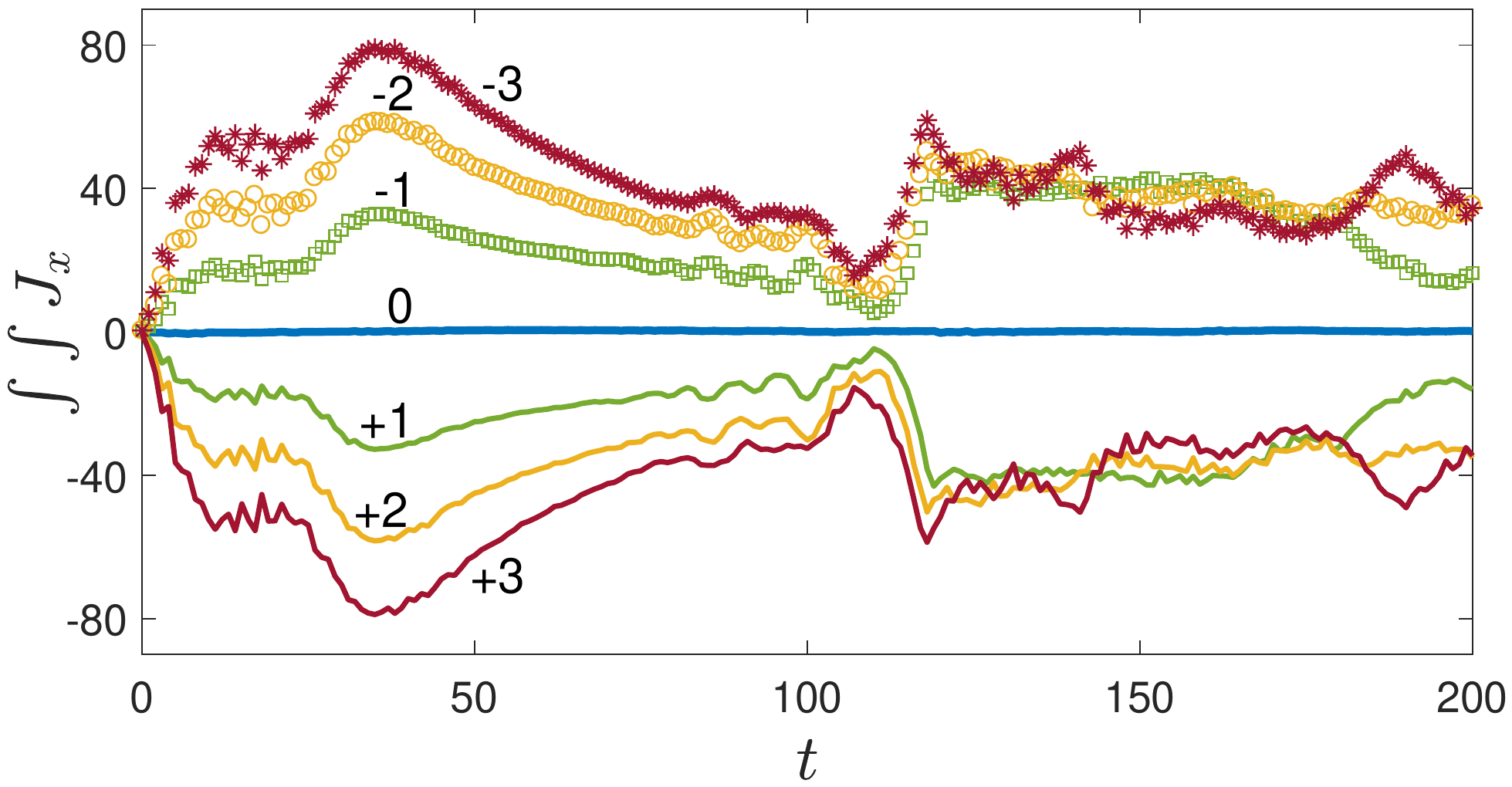}\\
\includegraphics[width=0.49\textwidth]{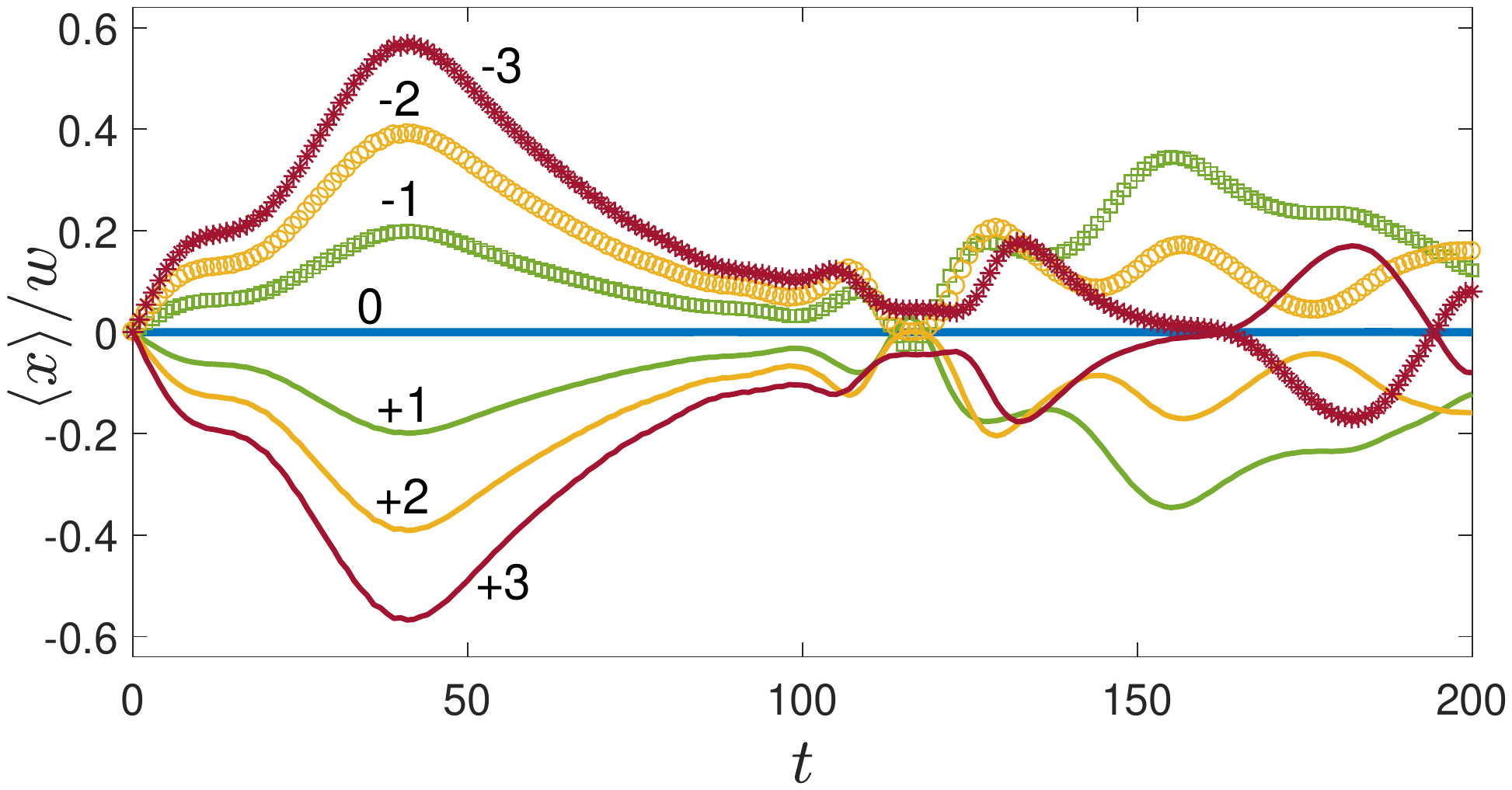}
\caption{Dynamics of the total number of atoms $N_{\text{tot}}$ (top), the net 
particle flux in $x$ direction $\int{dx}\int{dy}J_x$ (middle) and the $x$ component of 
the center of mass of atomic cloud $\langle x\rangle$ (bottom) inside the waveguide, for 
non-rotating, $\Omega=0$, and rotating cases with $\Omega=\pm1,\pm2,\pm3$. Value of $\Omega$ 
is indicated next to the corresponding curve in the middle and bottom panels. The ring potential 
and the waveguide have same width $w$ and depth $U_0=20$. Atom-atom interaction strength is 
$u=2\times10^{-4}$ and ABC potential strength is $V_0=20$.
\label{fig:omega}}
\end{figure}

Following the dynamics of the atoms inside the waveguide, we do not observe any reflection from the boundaries in $x$ direction. 
However, the distribution of the atoms in $x$ direction is not continuous in time due to fluctuation of the number of atoms which 
tunnel from the ring into the waveguide. For all $\Omega$, we observe very similar number of atoms inside the waveguide (top panel 
in Fig.~\ref{fig:omega}) indicating very similar tunneling rates (chemical potential $\mu$ in the ring has a very weak dependence 
on $\Omega$). Nevertheless, the current state inside the ring can be clearly read-out by looking at the imbalance between the right- 
and left-moving atoms as well as the center of mass position of the atomic density in the waveguide (middle and bottom panels of Fig.~\ref{fig:omega}). 
While the sign of these quantities reveals the direction of rotation inside the ring, their absolute value can be used to probe 
the magnitude of the winding number.

By inspection of Fig.~\ref{fig:omega}, we notice a marked dip (around $t\sim 110-120$) in all plotted quantities. 
Such a feature traces back to a specific collective phenomenon occurring in the ring condensate: The tunneling 
process results in perturbation of the density of the condensate. Such a perturbation decays in a pair of density 
modulations which counter-propagate along the ring with negligible dispersion; given the very small magnitude of 
the perturbations, the excitations can be of phononic-type. Analyzing our results further, we see that the dip occurs 
shortly after the time at which the density modulations recombine around the tunneling region. For the non-rotating 
case the counter-propagating excitations with same frequency move with same speed to meet again at the same point 
where they were produced. For the flowing currents, instead, the frequency of excitations, and therefore the velocity 
of the density perturbations, are affected by Doppler effect~\cite{SupMat,kumar2016minimally}, implying that the 
recombination point of the density perturbations is dragged along the superfluid current.

A simple Bogoliubov analysis of the idealized $1d$ ring condensate \cite{SupMat} gives results which quantitatively 
agree with the numerical outcome. In particular, the modulation of the density propagate as $\delta|\psi|^2 \propto \cos(q\phi\mp\omega^{\pm} t)$ 
where $\phi$ is angular coordinate along the ring and $q$ is the angular wavenumber of excitation. Here, $\omega^{\pm}=\omega_0\pm 2q\Omega/R^2$ 
are the enhanced and reduced frequencies (due to Doppler effect) of the two counter-propagating excitations and $\omega_0$ 
is the frequency of excitations in absence of rotation. The density perturbations produced by these excitations then 
travel with enhanced and reduced velocities $v^{\pm}=v_0\pm\Omega/R$, with $v_0$ being the velocity of density perturbations 
in absence of rotation, and reach their original place at times $T^{\pm}$, where $T^+<T^-$. We note that in Fig.~\ref{fig:omega}, 
for $\Omega=\pm2$ and $\pm3$, there are two dips in $N_{\text{tot}}$ around the time $t\approx120$ which indicate the 
time difference between the arrival of the fast and slow moving density perturbations at the tunneling point. This time 
difference has not been resolved in our numerical data for $\Omega=\pm1$ due to the finite length of the density perturbations 
and small velocity shift. However, the dip in $N_{tot}$ for this case is shallower and wider than the one of non-rotating case. 
It is remarkable that such a Doppler effect of the excitations implies clear signatures in all quantities measured in the 
waveguide. As a result of the Doppler shift, the meeting point of the density perturbations is dragged along the supercurrent 
and when the perturbations meet around the tunneling region for the first time at $t=\pi R/v_0$ there is a Sagnac phase-shift 
of $k_q\omega_s{\cal{A}}/v_0$~\cite{anderson1994sagnac}. Here $k_q=q/R$ is the wavenumber of the excitations, ${\cal{A}}$ is 
the area of the circle and $\omega_s=2\Omega/R^2$ is the angular velocity of the supercurrent.

After the density perturbations reach back to the tunneling point the atoms' distribution in waveguide becomes more complicated: 
The rotating states are still detectable from non-rotating state through the asymmetry in the net particle flux in the waveguide 
given by $\int{dx}\int{dy}J_x$; the states with different winding numbers, however, seem not to be distinguishable through the 
quantities shown in Fig.~\ref{fig:omega}. Such time depends on the interaction strength through the group velocity of the rotating 
density perturbation (see \cite{SupMat} for details). Therefore, with weaker interactions, the maximum time for which the 
rotating states are well-differentiated from each other is extended. On the other hand, the interaction reshuffles the configuration 
of the energy levels (through the chemical potential of ring condensate), affecting in turn the ring-guide tunneling rate.

\begin{figure}[t!]
\includegraphics[width=0.235\textwidth]{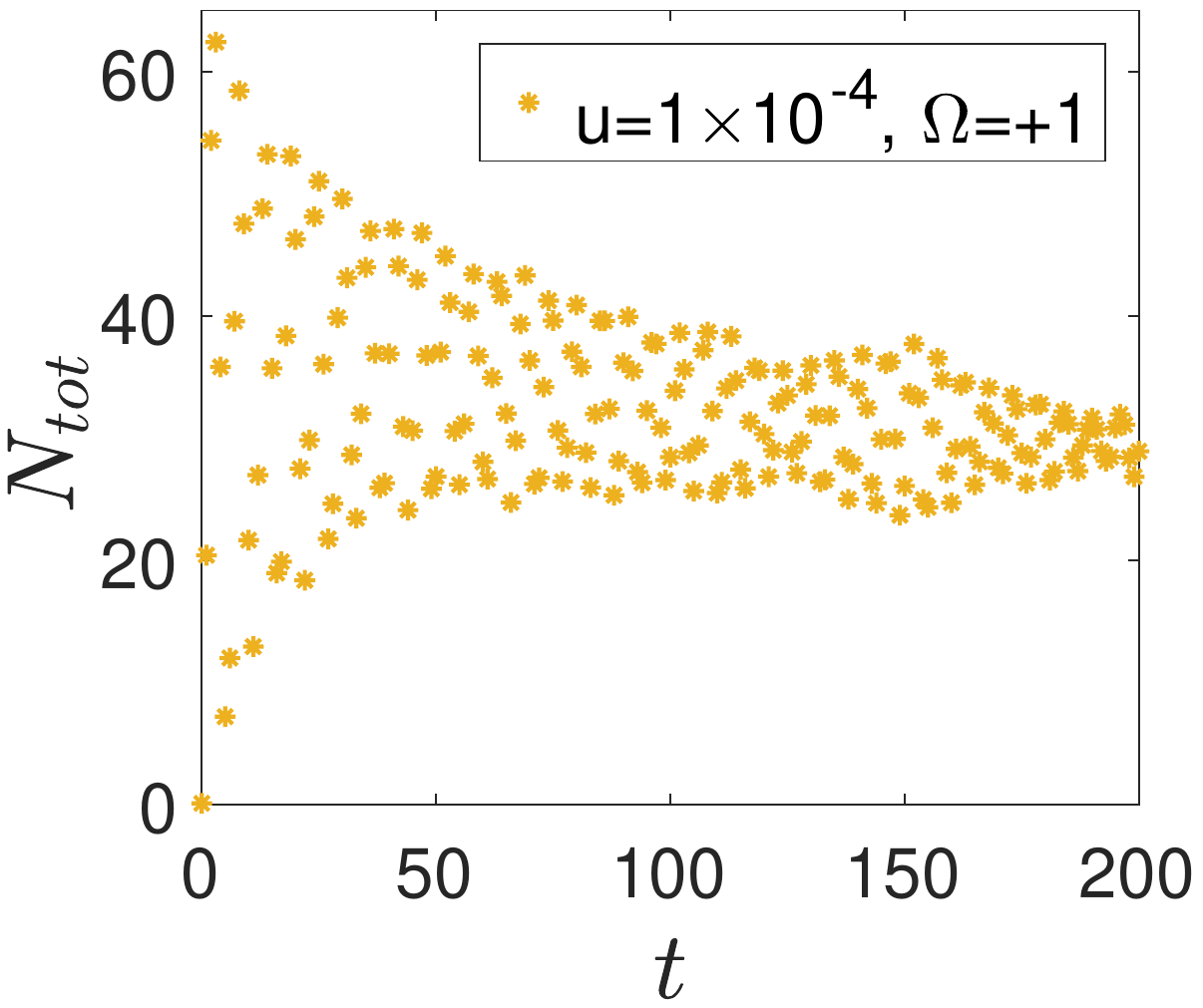}
\includegraphics[width=0.235\textwidth]{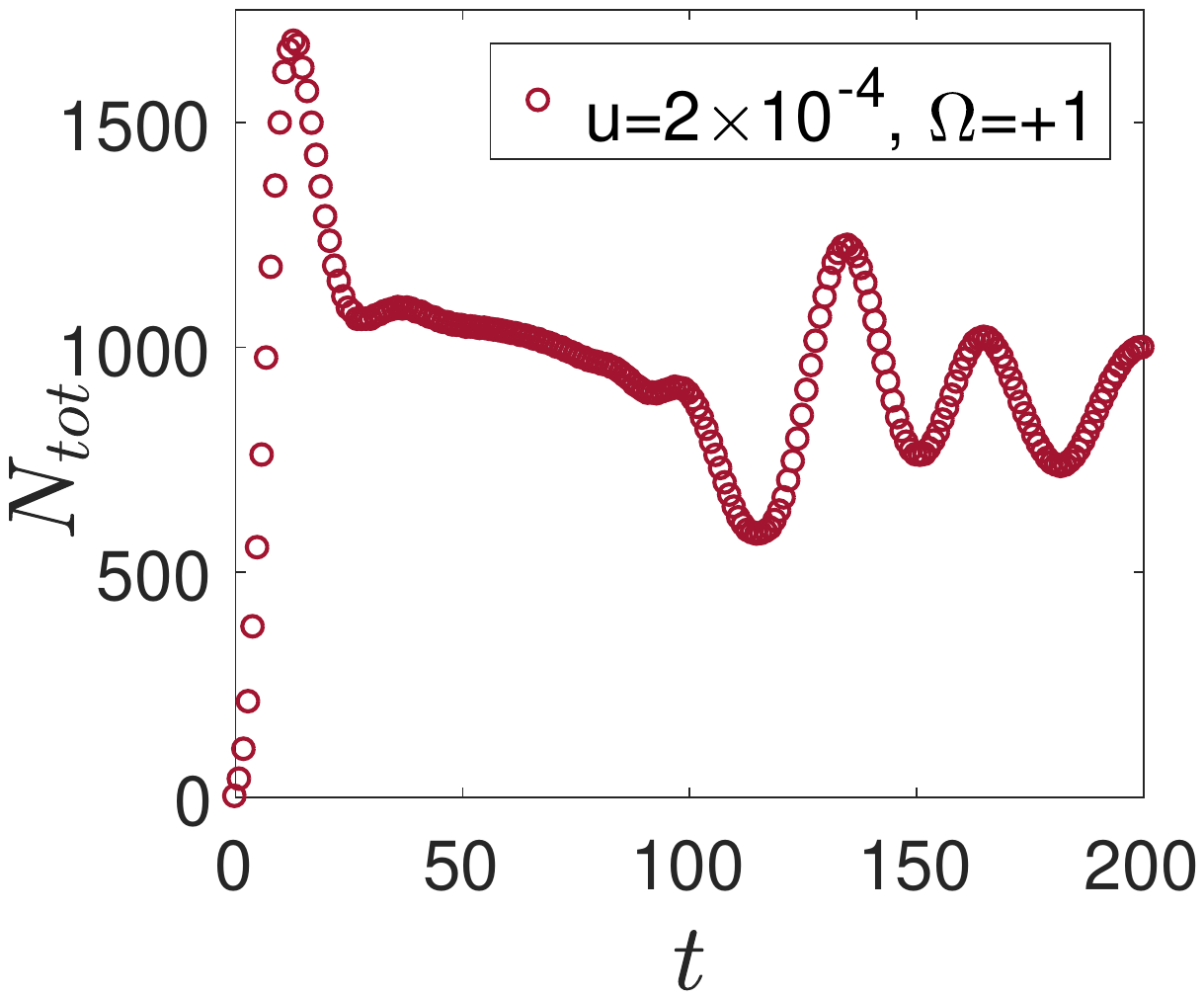}\\
\includegraphics[width=0.235\textwidth]{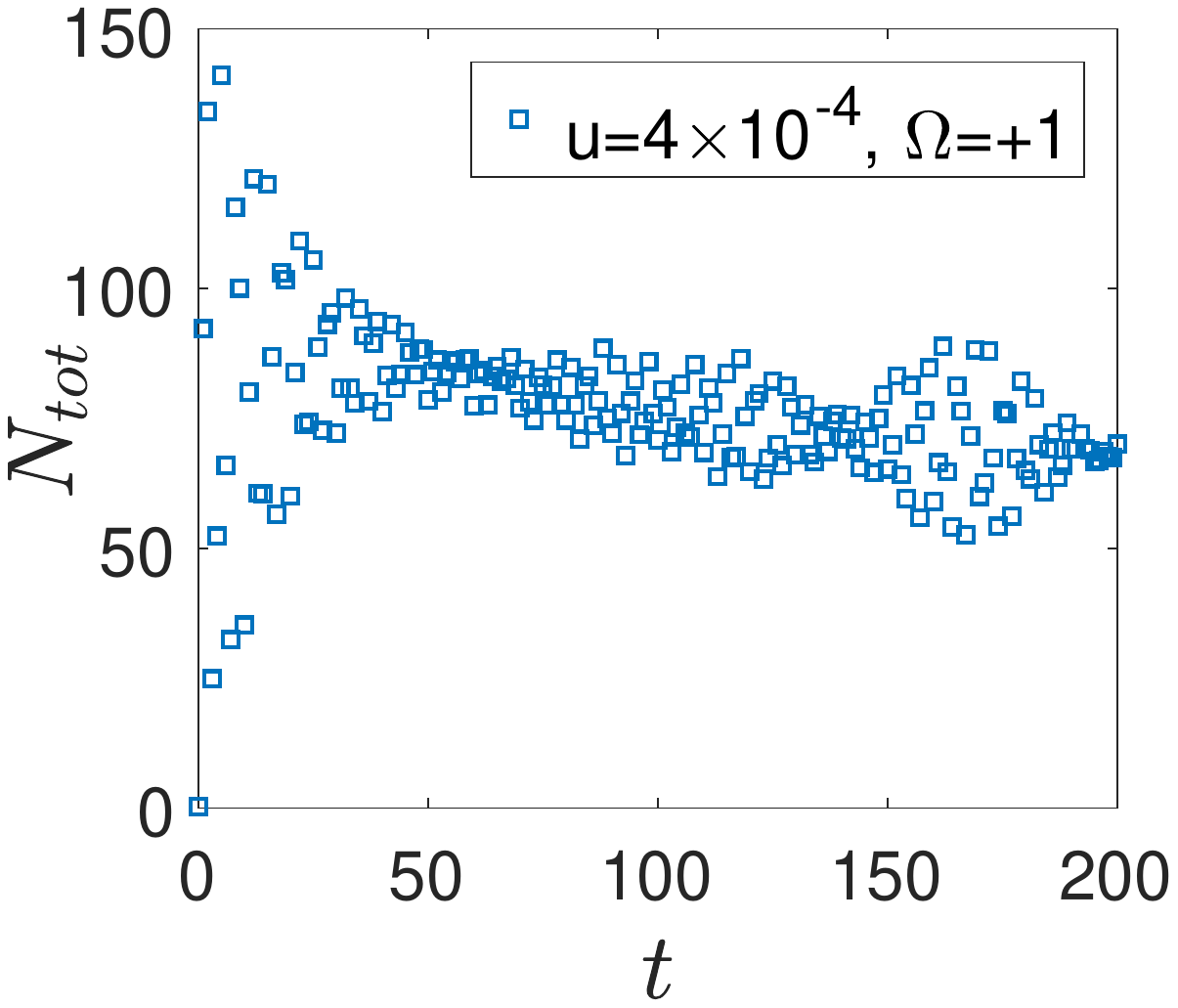}\\
\includegraphics[width=0.5\textwidth]{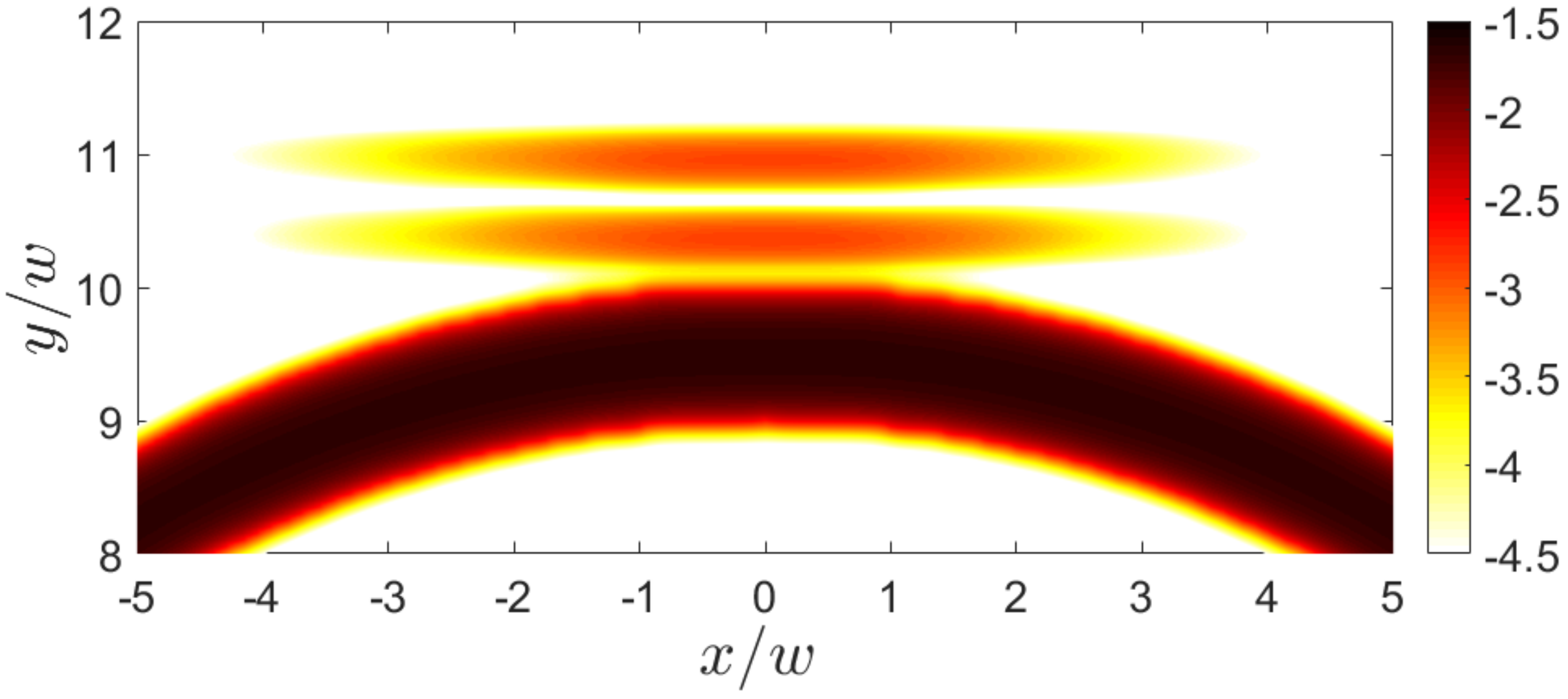}
\caption{Three top panels: the total number of particles inside the waveguide in time, for the rotating 
states with $\Omega=+1$ and three different values of the atom-atom interaction strength $u$. The 
ring potential and the waveguide have same width $w$ and depth $U_0=20$ and the state in the ring 
with $u=2\times10^{-4}$ (top-right panel) is in better resonance with the energy levels of the waveguide 
compared to other cases. ABC potential strength is $V_0=20$. Bottom panel: the logarithm of the 
atomic density in the waveguide and top side of the ring in vicinity of waveguide at time $t=13$ 
for the case with $\Omega=+1$ and $u=2\times10^{-4}$. The density profile in $y$ direction inside 
the waveguide indicates that the second discrete level is occupied as expected.
\label{fig:resonance1}}
\end{figure}

\begin{figure}[t!]
\includegraphics[width=0.235\textwidth]{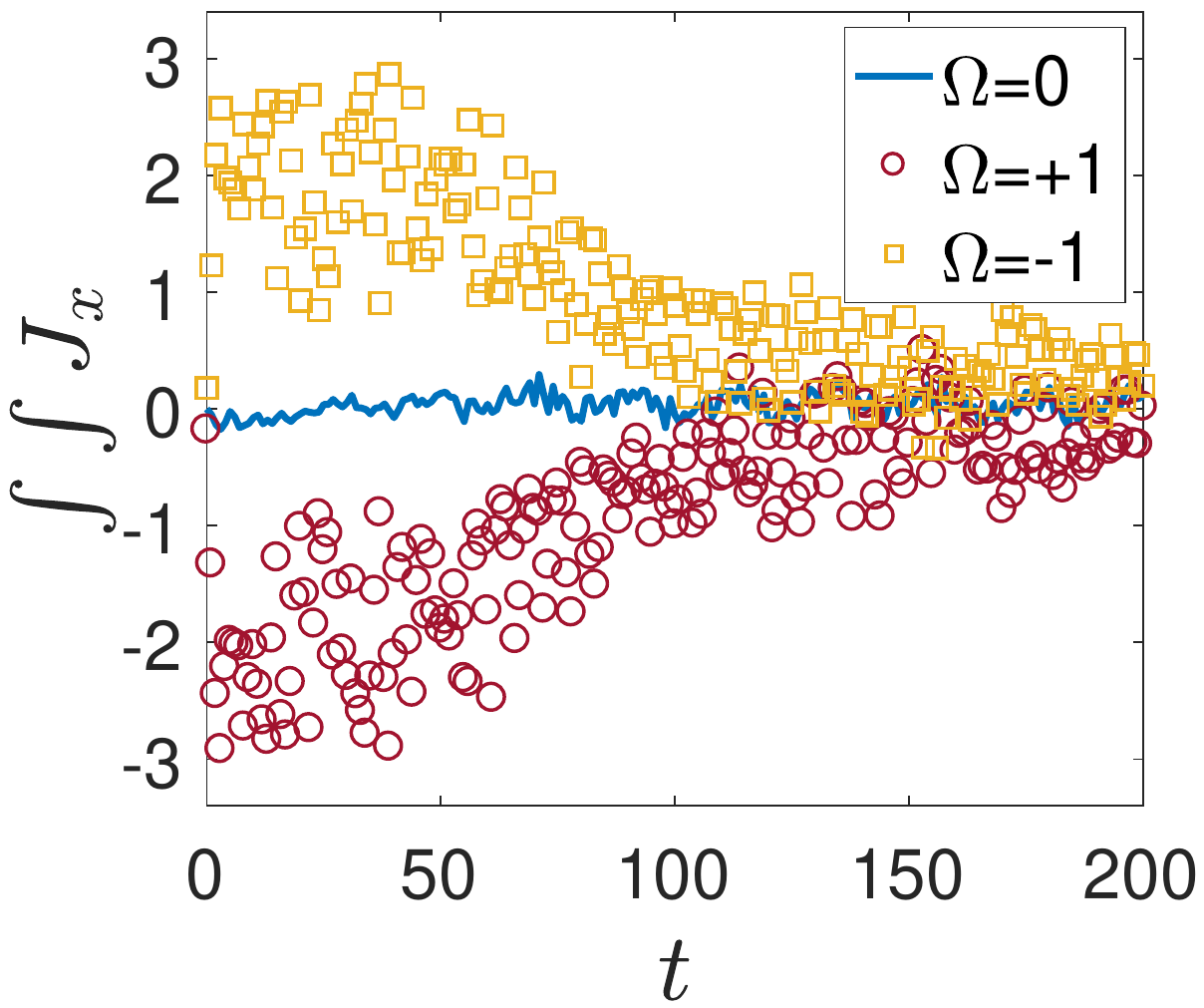}
\includegraphics[width=0.235\textwidth]{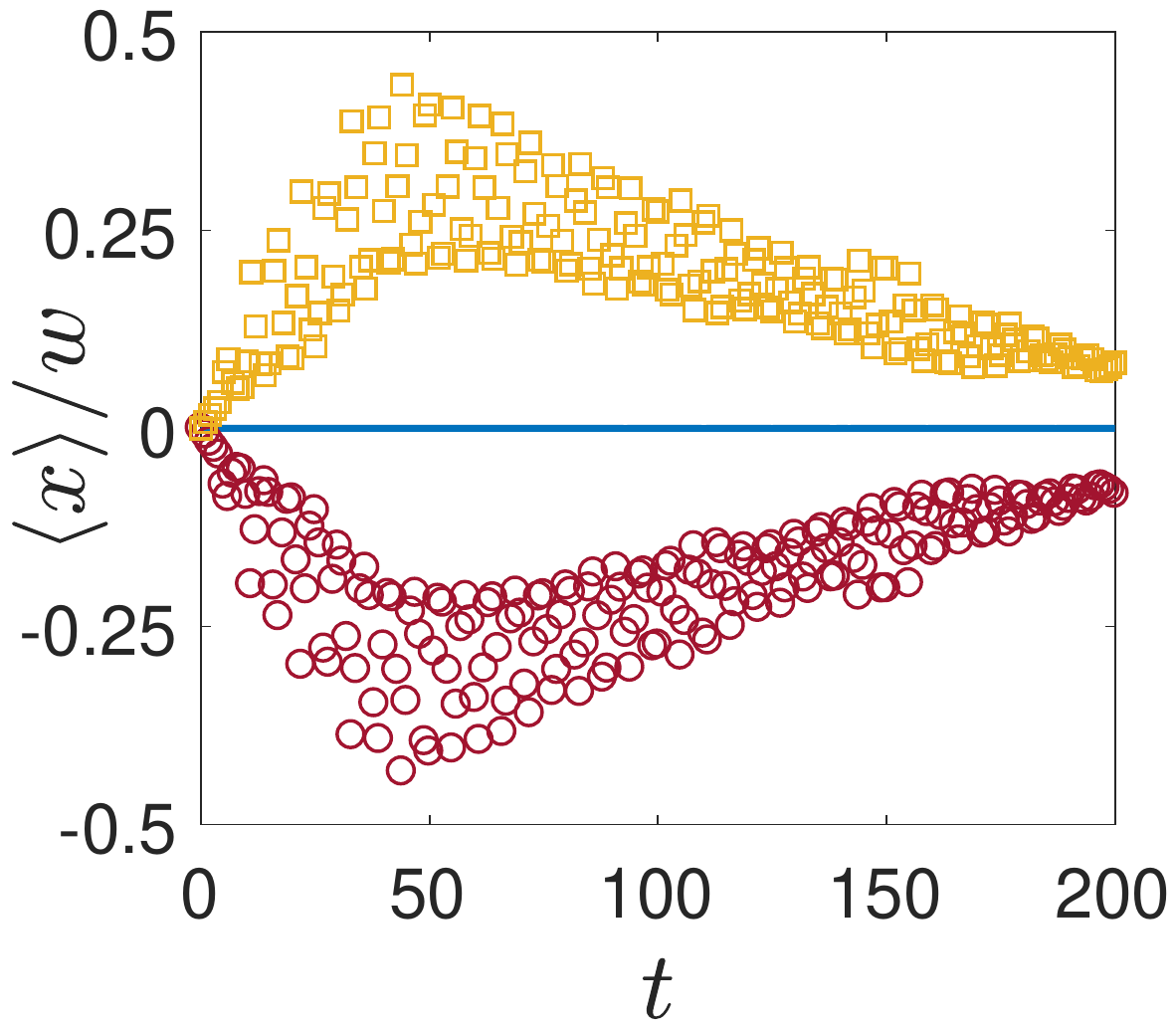}\\
\includegraphics[width=0.235\textwidth]{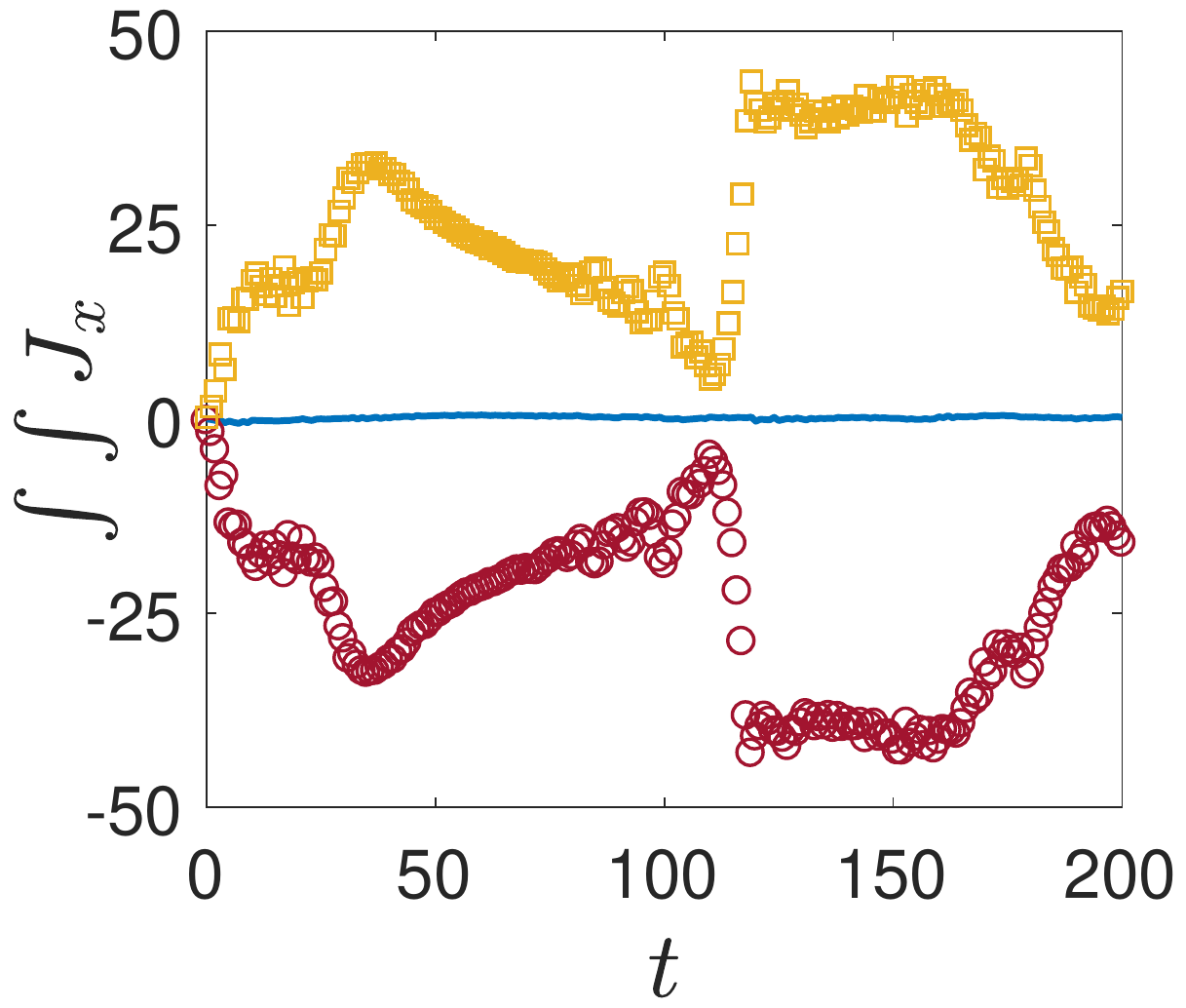}
\includegraphics[width=0.235\textwidth]{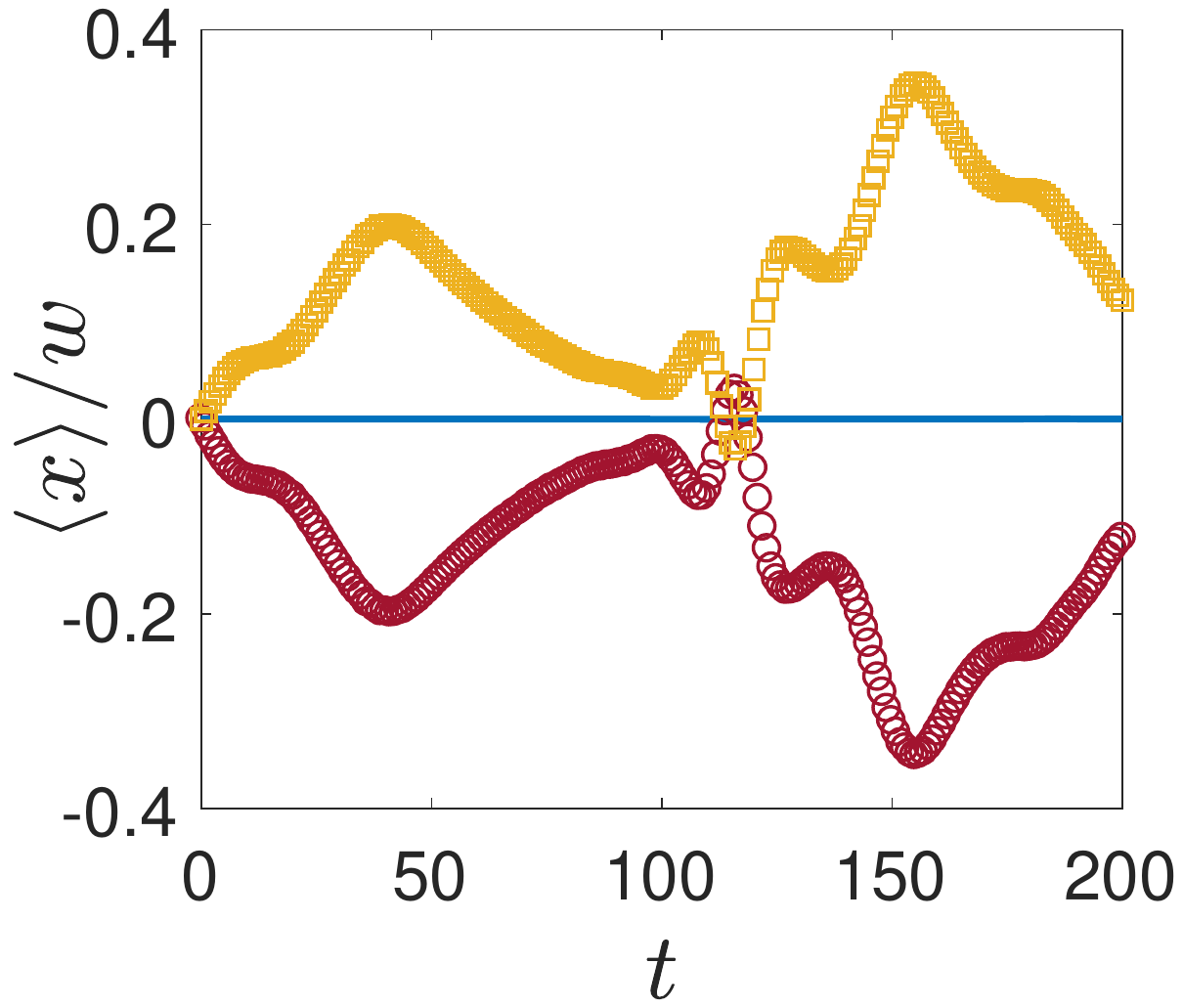}\\
\includegraphics[width=0.235\textwidth]{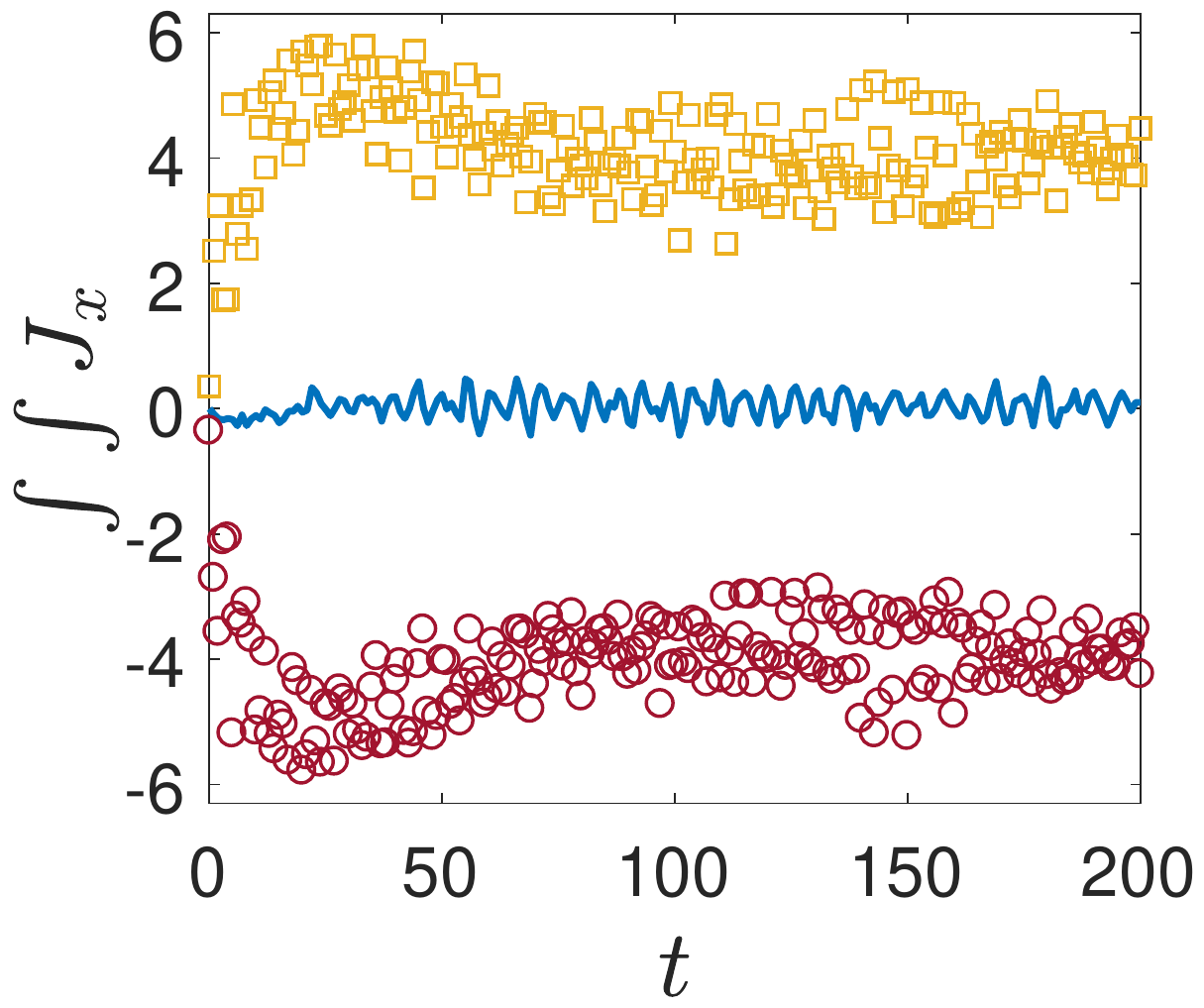}
\includegraphics[width=0.235\textwidth]{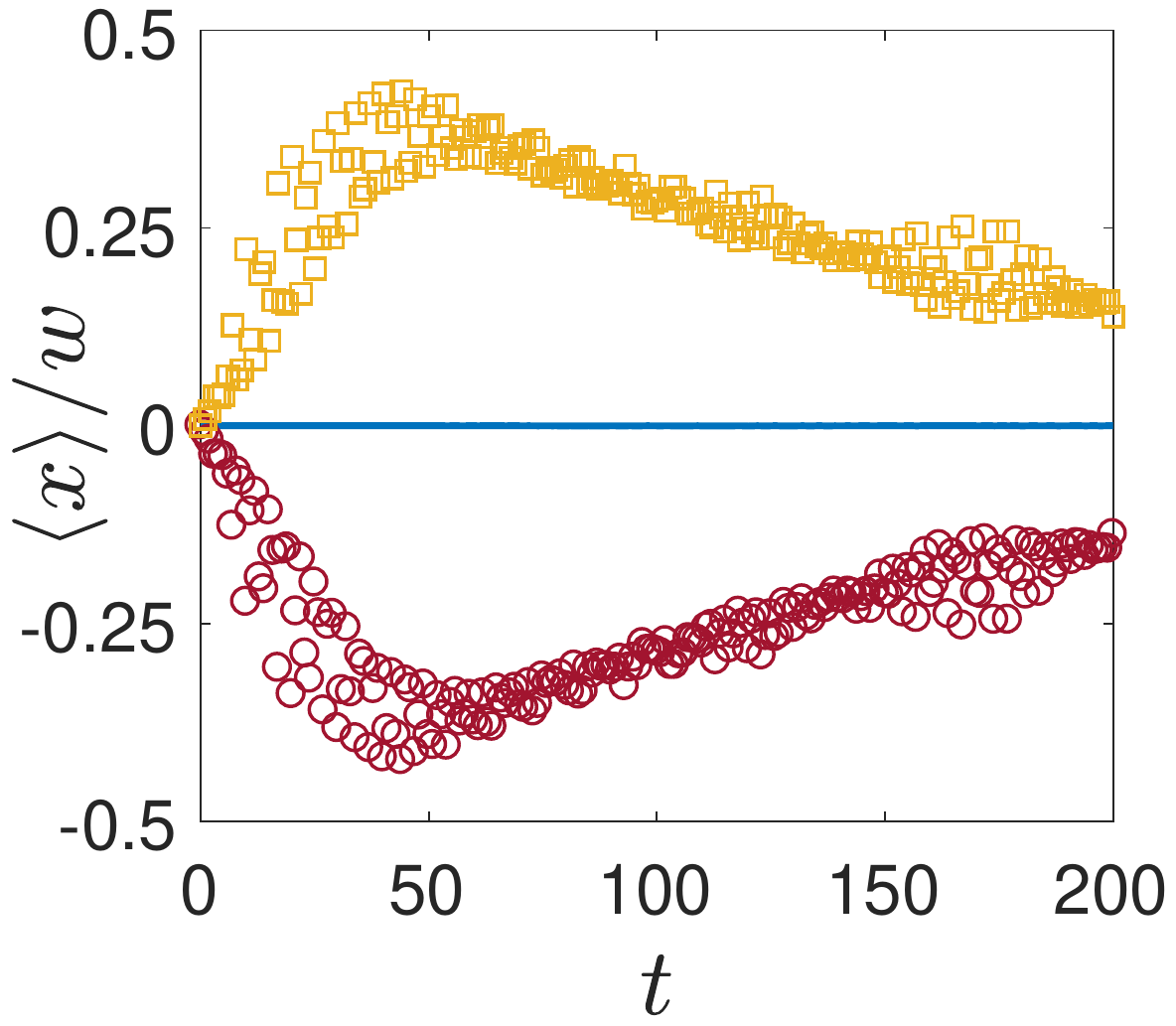}
\caption{Dynamics of the net particle flux in $x$ direction $\int{dx}\int{dy}J_x$ 
(left panels) and the $x$ component of the center of mass of atomic cloud $\langle x\rangle$ 
(right panels) inside the waveguide, for three different values of the atom-atom interaction 
strength $u=1\times10^{-4},~2\times10^{-4}$ and $4\times10^{-4}$ from top to bottom. Other 
parameters are the same as those in Fig.~\ref{fig:resonance1}. The blue/solid lines correspond 
to the case with $\Omega=0$, the red/dark circles to $\Omega=+1$ and the yellow/light squares to 
$\Omega=-1$. 
\label{fig:int}}
\end{figure}

Table \ref{tab:one} summarizes the difference between the chemical potential of atoms in the ring and three lowest discrete energy 
levels in the waveguide for three different values of atom-atom interaction strength $u$. Three top panels of the Fig.~\ref{fig:resonance1} 
show the total number of particles inside the waveguide for the rotating states with $\Omega= 1$ and the three different values of 
the interaction strength displayed in the Table \ref{tab:one}. We observe that the highest resonance case, with $u=2\times10^{-4}$, 
corresponding to the highest tunneling rate, is characterized by a `clean' time dependence. For larger detuning, in contrast the 
tunneling is much more erratic. This behavior suggests that, while the off-resonant ring-guide tunneling involves different frequencies, 
the near-resonant tunneling involves mostly a single level (the resonant one). Indeed, we see that it is the second discrete state 
(due to confinement in $y$ direction) to be involved in this case (bottom panel of Fig.~\ref{fig:resonance1}). We note that, despite 
the small number of atoms in the waveguide for the off-resonant cases, the asymmetry due to rotation is still observed in the quantities 
plotted in Fig.~\ref{fig:int}.
\begin{table}
  \begin{center}
    \begin{tabular}{|c|c|c|c|c|}
     \hline
      \textbf{$u$} & \textbf{$\mu$} & \textbf{$\mu-E_1$} & \textbf{$\mu-E_2$} & \textbf{$\mu-E_3$}\\
      \hline
      $1\times10^{-4}$ & $-18.11$ & $1.13$ & $-1.15$ & $-4.89$\\
      $2\times10^{-4}$ & $-17.07$ & $2.17$ & $-0.11$ & $-3.85$\\
      $4\times10^{-4}$ & $-15.12$ & $4.12$ & $1.84$ & $-1.9$\\
     \hline
    \end{tabular}
		\caption{The difference between the chemical potential of atoms in the ring 
		$\mu$ and three lowest energy levels inside the waveguide for three different 
		values of the atom-atom interaction strength $u$. All other parameters are the 
		same as those in Fig.~\ref{fig:omega}. Rotation does not change the first two 
		decimal digits of the $\mu$ given here. The case with $u=2\times10^{-4}$ has 
		the highest resonance with the second level inside the waveguide.
    \label{tab:one}}
  \end{center}
\end{table}
The resonant cases correspond to a large number of atoms tunneling from the ring to the waveguide, 
causing a substantial decrease of the density in the ring condensate. Indeed, the resonance condition 
can be controlled by tuning the waveguide's parameters.

\section{Notes on Experimental Implementation}
Here we briefly discuss the feasibility of the proposed system in the experiment.
 
First we would like to mention that the step-function potentials are considered in this work for convenience 
in order to make it easier to tune the distance between the ring potential and the waveguide. Even though 
with the use of new technologies, such as SLM, fabrication of versatile forms of optical potentials has been 
made possible, we emphasis that what actually matters is the tunneling rate between the ring potential and 
the waveguide. Therefore, depending on the experimental setup, either the distance or the resonance between 
the energy levels of the two potentials can be used to control the tunneling rate. The resonance can also be 
controlled by either the depth or width of each potential. One could imagine that tuning and changing the 
geometrical parameters of the rectilinear waveguide is more convenient compared to changing the parameters of 
the ring potential.

Second point to consider is the ratio of the ring's radius $R$ with respect to its width $w$. In this work we 
have considered a rather tight ring potential such that, for all values of the gauge field which are used, only 
one winding number is permitted in the ring. In other words, the winding number does not change from the inner 
radius to the outer radius of the ring. This condition is imposed mainly to avoid complications in numerical 
simulations. The aim has been to avoid excitation of unwanted states with higher winding numbers. Depending on 
the method used in the experiment to bring the atoms into rotation the $R/w$ ratio may not be of any concern.

As for the measurement time restrictions, if we consider $^{87}Rb$ atoms, for instance, in a ring potential with 
a width $w=1~\mu m$, the unit of time becomes $\omega^{-1}\approx277.45~\mu s$. This means that the measurement 
must be performed within a time of $t\approx100~\omega^{-1}\approx27.7~ms$. 
We have also worked with dimensionless atom-atom interaction strengths $\tilde{u}=1,~2,~4\times10^{-4}$ which 
are equivalent to scaled scattering lengths $a_s/\delta=4,~8,~16\times10^{-6}$ for $^{87}Rb$ atoms. In a rough 
approximation the $3$D-to-$2$D scaling factor $\delta$ is equal to the size of the system in the transverse ($z$) 
dimension~\cite{petrov2001}. Therefore, for a system with tight confinement in third dimension these values of 
$\tilde{u}$ represent very weak interactions.  

\section{Conclusion}
%
We provided a numerical analysis of the quench dynamics of a specific atomic circuit made of a ring-shaped bosonic 
condensate coupled with a rectilinear waveguide of finite length. We demonstrated that both magnitude and the direction 
of the current flowing through the ring can be detected through the inspection of the very small number of atoms 
tunneling from the ring into the waveguide. The protocol we conceived is minimally destructive on the ring condensate 
and allows to carry-out the measurements of the flowing states in a virtually continuous way while the ring operates. 
Interestingly enough, we find that the dynamics in the circuit is characterized by a peculiar effect: the depletion of 
the condensate density, caused by the ring condensate-waveguide tunneling, decays into a pair of phonon-type excitations. 
These excitations meet again, after they have traveled along the loop, in a position that is fixed by the Doppler effect 
induced by the persistent current and characterized by a Sagnac phase shift. Such effect plays a key role for the read-out 
protocols. At the same time, it could be exploited to access the predictions implied in the quasi-particles decay in Bose 
condensates~\cite{Beliaev1957,Giorgini1998,Glazman2010,Matveev2016,ozeri2005colloquium}. In particular the crossover in 
the spatial dimension (from $3d$ down to $1d$) and interaction can be explored. In addition, by playing with the ring-guide 
coupling, one could produce density excitations of more substantial magnitudes (soliton-like), with different pair formation 
mechanism~\cite{malomed1989,Hulet2014}. We believe that our work will play an instrumental role for the diagnostics of cold-atoms 
systems with non-trivial winding numbers. We have also shown that fundamental physics is implied in the dynamics of the system. 
Finally, our circuit provides the basis for a new architecture of rotation sensors. 

\section{Acknowledgments}
%
We would like to acknowledge fruitful discussions with  B. Gr\'{e}maud, T. Haug and C. Miniatura. 
This research is supported by the National Research Foundation, Prime Minister's Office, Singapore 
and the Ministry of Education-Singapore, under the Research Centres of Excellence programme and 
Academic Research Fund Tier 2 (Grant No. MOE2015-T2-1-101). The computational work for this article was 
mainly performed on resources of the National Supercomputing Centre, Singapore (https://www.nscc.sg).
The Grenoble LANEF framework (ANR-10-LABX-51-01) is acknowledged for its support with mutualized
infrastructure.

\begin{appendix}
\counterwithin{figure}{section}
%
\section{Numerical method}
\label{numerics}
To compute the dynamics of the system governed by equation (1) of the main text, in real or imaginary time, 
we use a generalized version of the Split-Step Method developed in~\cite{bao2006efficient}, where a gauge field of the 
form $\vec{A}(x,y)=A_x(y)\hat{x}+A_y(x)\hat{y}$ is considered. This method covers the gauge field that we have used 
in this work as long as $B$ is constant everywhere. For the numerical results presented in this paper, we first compute 
the ground state of the ring potential, with different values of magnetic field $B$, by integrating (1) in 
imaginary time. In this case $V(\vec{r})=V_r(x,y)$ while $V_g(x,y)=0$. For the real time dynamics, beginning with the 
obtained ground state, we turn on the waveguide potential by considering $V(\vec{r})=V_r(x,y)+V_g(x,y)$ while, at the 
same time, setting the gauge field to zero in order to avoid any effect of gauge field on the dynamics of the atoms 
which tunnel from the ring to the waveguide.

\section{Excitations in presence of supercurrent}
\label{excitations}
As it is mentioned in the main text, the weak tunneling of the atoms from ring to the waveguide produces excitations 
in the wavefunction of the BEC inside the ring. To better understand the dynamics of these excitations in presence of 
the supercurrent, we present some calculations by applying Bogoliubov excitations on the condensate. Since the 
density modulations are very small and appear on the tip of the density in the ring, we consider a one-dimensional 
system, essentially a ring with a fixed radius $R$ and azimuthal angle $\phi$, for simplicity. The ground state 
wavefunction of $N$ atoms on such a ring will have a form of $\psi_0(\phi)=\sqrt{n}e^{i\Phi(\phi)}$ with $n=N/(2\pi R)$ 
being the density of the atoms and $\Phi(\phi)$ the phase of the wavefunction. For a non-rotating condensate $\Phi(\phi)=const.$, 
while for a rotating condensate the gradient of this phase is proportional to the supercurrent velocity $v_s$: 
$\partial_{\phi}\Phi(\phi)=mRv_s/\hbar$. This wavefunction satisfies the time-independent GPE
\begin{eqnarray}
\label{eq:ground}
\mu\psi_0(\phi)=-\frac{\hbar^2}{2mR^2}\partial^2_{\phi}\psi_0(\phi)+U_0|\psi(\phi)|^2\psi_0(\phi)
\end{eqnarray}
with chemical potential 
\begin{eqnarray}
\label{eq:mu}
\begin{aligned}
\mu=\frac{\hbar^2}{2mR^2}\left(\partial_{\phi}\Phi\right)^2+nU_0=\frac{1}{2}mv_s^2+nU_0,
\end{aligned}
\end{eqnarray}
where we have assumed a constant supercurrent velocity, meaning that $\partial_{\phi}\Phi=const.$ and $\partial^2_{\phi}\Phi=0$. 
Therefore, the time-dependent wavefunction of the ground state reads $\psi_0(\phi,t)=e^{-i\mu t/\hbar}\sqrt{n}e^{i\Phi(\phi)}$ 
which satisfies the time-dependent GPE:
\begin{eqnarray}
\label{eq:ground-t}
\begin{aligned}
&i\hbar\partial_t\psi_0(\phi,t)=\\
&-\frac{\hbar^2}{2mR^2}\partial^2_{\phi}\psi_0(\phi,t)+U_0|\psi(\phi,t)|^2\psi_0(\phi,t).
\end{aligned}
\end{eqnarray}

We consider Bogoliubov excitations on top of the ground state and introduce the perturbed wavefunction 
$\psi(\phi,t)=\psi_0(\phi,t)+\delta\psi$. Assuming that the perturbed wavefunction $\psi(\phi,t)$ also 
satisfies GPE, and keeping only the terms which are linear in $\delta\psi$, the linearized dynamical 
equation reads: 
\begin{eqnarray}
\label{eq:perturb-dyn}
\begin{aligned}
&i\hbar\partial_t\delta\psi=\\
&-\frac{\hbar^2}{2mR^2}\partial^2_{\phi}\delta\psi+2nU_0\delta\psi+U_0\psi_0^2(\phi,t)\delta\psi^*.
\end{aligned}
\end{eqnarray}
By inserting excitations of the form
\begin{eqnarray}
\label{eq:delpsi1}
\delta\psi=e^{-i\mu t/\hbar}\left(u(\phi)e^{-i\omega t}+v^*(\phi)e^{i\omega t}\right)
\end{eqnarray}
into (\ref{eq:perturb-dyn}), and using the value of the chemical potential given in (\ref{eq:mu}) we find 
\begin{eqnarray}
\label{eq:uv}
\begin{aligned}
&\hbar\omega u=\\
&\left(\left(-\frac{\hbar^2}{2mR^2}\right)\left(\partial^2_{\phi}+(\partial_{\phi}\Phi)^2\right)+nU_0\right)u+nU_0e^{2i\Phi}v\\
-&\hbar\omega v=\\
&\left(\left(-\frac{\hbar^2}{2mR^2}\right)\left(\partial^2_{\phi}+(\partial_{\phi}\Phi)^2\right)+nU_0\right)v+nU_0e^{-2i\Phi}u.
\end{aligned}
\end{eqnarray}

With a steady supercurrent ($\partial_{\phi}\Phi=const.$) the equations (\ref{eq:uv}) have solutions
\begin{eqnarray}
\label{eq:uvdef}
\begin{aligned}
&u=Ae^{i(q\phi+\Phi)}\\
&v=Be^{i(q\phi-\Phi)}.
\end{aligned}
\end{eqnarray}
The quantities $q$, $A$ and $B$ are related by
\begin{eqnarray}
\label{eq:AB}
\begin{aligned}
&\hbar\omega=\frac{\hbar^2}{2mR^2}\left(q^2+2qc\right)+nU_0\left(1+\frac{A}{B}\right)\\
-&\hbar\omega=\frac{\hbar^2}{2mR^2}\left(q^2-2qc\right)+nU_0\left(1+\frac{B}{A}\right),
\end{aligned}
\end{eqnarray}
where $c=\partial_{\phi}\Phi=mRv_s/\hbar$ is the constant phase gradient of the ground state. 
One can solve (\ref{eq:AB}) for the dispersion relation of the excitations:
\begin{eqnarray}
\label{eq:omega}
\hbar\omega=\frac{\hbar^2qc}{mR^2}\pm\hbar\omega_0\doteq \pm\omega^{\pm},
\end{eqnarray}
where $\hbar\omega_0=\sqrt{\epsilon_q(\epsilon_q+2nU_0)}$ and $\epsilon_q=\hbar^2q^2/(2mR^2)$. In absence of supercurrent ($c=0$) excitations 
have a single frequency $\omega_0$. However, in presence of the supercurrent the frequency is shifted by $\hbar qc/(mR^2)=qv_s/R$.

Assuming that $\Phi(\phi)=c\phi$ and substituting (\ref{eq:uvdef}) into (\ref{eq:delpsi1}) results in
\begin{eqnarray}
\label{eq:delpsi2}
\begin{aligned}
&\delta\psi=\\
&e^{-i\mu t/\hbar}\left(Ae^{-i\omega t+i(q+c)\phi}+B^*e^{i\omega t-i(q-c)\phi}\right)
\end{aligned}
\end{eqnarray}
and therefore, the linearized perturbation of the density $\delta|\psi|^2=|\psi|^2-|\psi_0|^2=\psi_0\delta\psi^*+\psi_0^*\delta\psi$ reads as
\begin{eqnarray}
\label{eq:deldensity}
\begin{aligned}
\delta|\psi|^2&=2\sqrt{n} \Re\left((A+B)e^{-i\omega t+iq\phi}\right)\\
&=\sqrt{n}(A+B)\cos(q\phi-\omega t)~;~A,B\in{\rm I\!R}
\end{aligned}
\end{eqnarray}
which has the form of a sound wave with 
\begin{eqnarray}
\label{eq:amp1}
\begin{aligned}
A+B=
\begin{cases}
\left(\frac{\hbar\omega_0-\epsilon_q}{nU_0}\right)B &; \omega=\omega^+\\
-\left(\frac{\hbar\omega_0+\epsilon_q}{nU_0}\right)B &; \omega=-\omega^-
\end{cases}
\end{aligned}
\end{eqnarray}
The excitation with frequency $\omega=\omega^+$ produces density perturbations of the form $\cos(q\phi-\omega^+t)$, while the other 
one with $\omega=-\omega^-$ causes perturbations with $\cos(-q\phi-\omega^-t)$ profile. Therefore, the density perturbation which 
moves along the supercurrent has higher velocity and smaller amplitude and is the result of the excitation with enhanced frequency while the one in the 
opposite direction has smaller velocity with larger amplitude and is caused by excitations with lowered frequency. The group velocity of these density 
perturbations are given by 
\begin{eqnarray}
\label{eq:velocity}
v^{\pm}=R\partial_q\omega^{\pm}=v_0\pm\hbar c/(mR),
\end{eqnarray}
where $v_0=(\hbar^2q^2/(2m^2R^2)+c_s^2)/\sqrt{\hbar^2q^2/(4m^2R^2)+c_s^2}$ is the velocity in absence of supercurrent 
and depends on the wavenumber of the excitations $q$ as well as the sound velocity $c_s=\sqrt{nU_0/m}$. However, the 
shift in the velocity $\pm\hbar c/(mR)$ only depends on the gradient of the phase $c$ due to supercurrent. Dependence 
on radius $R$ appears here only because we have considered a circle and gradient is defined in $\phi$ direction (see (\ref{eq:ground})). 
For the case of a straight line, $\phi\rightarrow x$ and $R\rightarrow 1$.

In summary, due to Doppler effect, there is a phase shift of $\pm\hbar qct/(mR^2)=\pm qv_st/R$ for the two counter propagating density 
perturbations. With a simple calculation one can show that the excitations meet for the first time at $t=2\pi R/v_0$ and therefore the 
resulting Sagnac phase-shift is equal to $2k_q{\cal{A}}\omega_s/v_0$ where ${\cal{A}}=\pi R^2$ is the area of the circle, $\omega_s=v_s/R$ 
is the angular velocity of the supercurrent and $k_q=q/R$ is the linear wavenumber of the excitations.

For the system studied in the main text, the gradient of the phase in $\phi$ direction is equivalent to the winding number $\Omega$. Using the 
dimensionless quantities of the main text, one can rewrite the frequencies of the excitations as $\omega^{\pm}=\omega_0\pm 2q\Omega/R^2$. 
Therefore at time $t=\pi R/v_0$, when the two density modulations meet for the first time, the corresponding Sagnac phase-shift is equal to 
$k_q\omega_s{\cal{A}}/v_0$. The dimensionless sound and group velocities read as $c_s=\sqrt{Nu|\psi|^2/2}$, $v_0=(q^2/(2R^2)+c_s^2)/\sqrt{q^2/(4R^2)+c_s^2}$, 
and $v^{\pm}=v_0\pm \Omega/R$. Therefore on a circle with radius $R$, one would expect the fast and slow excitations to make a full circle and 
return to the their production point at times $T^{\pm}=\pi R/v^{\pm}=(1/T_0\pm\Omega/(\pi R^2))^{-1}$, with $T_0=\pi R/v_0$ being the returning 
time in absence of supercurrent.

\begin{widetext}
\begin{figure*}[t!]
\includegraphics[width=0.25\textwidth]{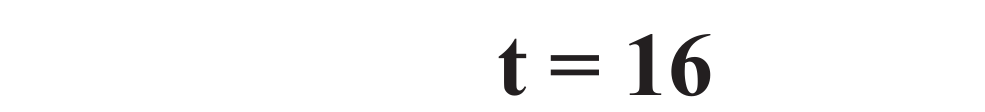}
\includegraphics[width=0.25\textwidth]{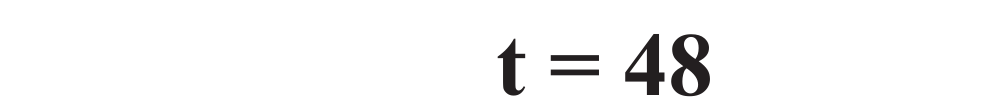}
\includegraphics[width=0.25\textwidth]{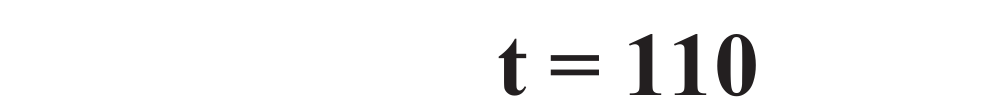}\\
\includegraphics[width=0.05\textwidth]{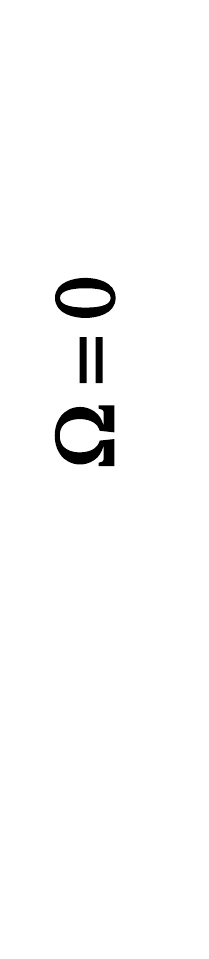}
\includegraphics[width=0.25\textwidth]{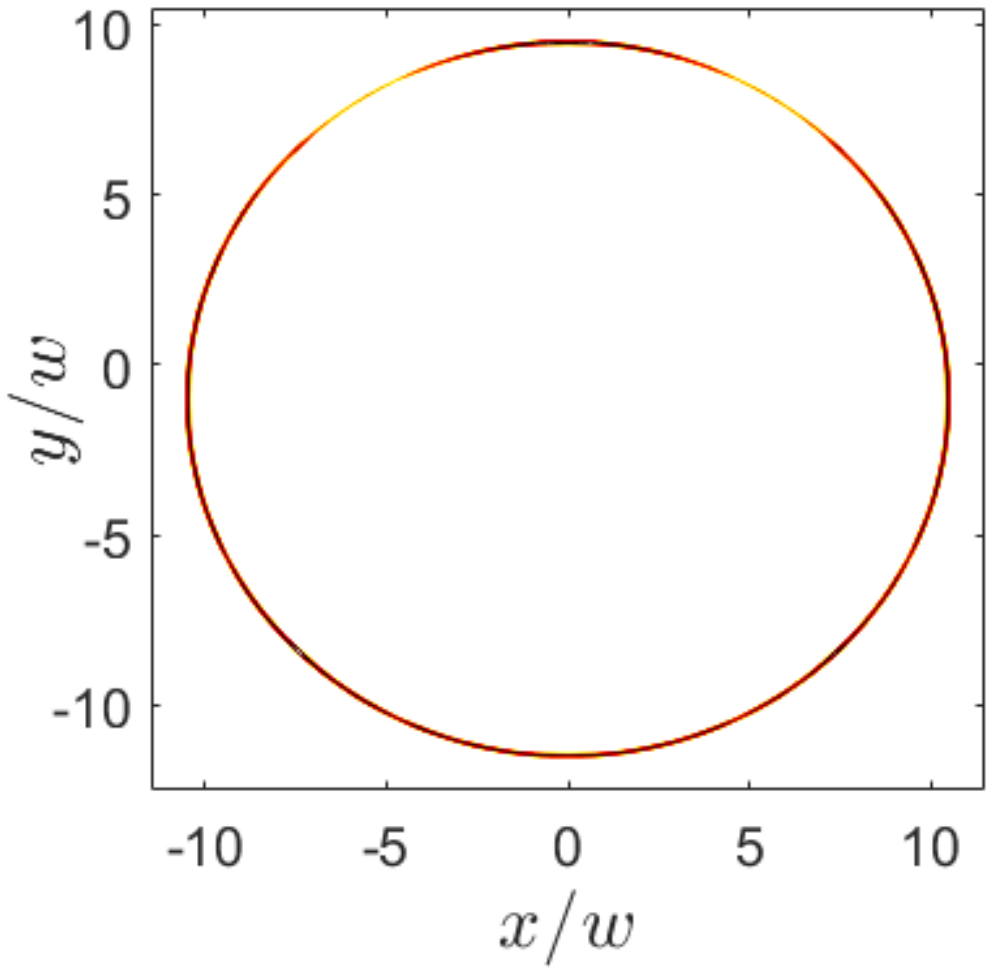}
\includegraphics[width=0.25\textwidth]{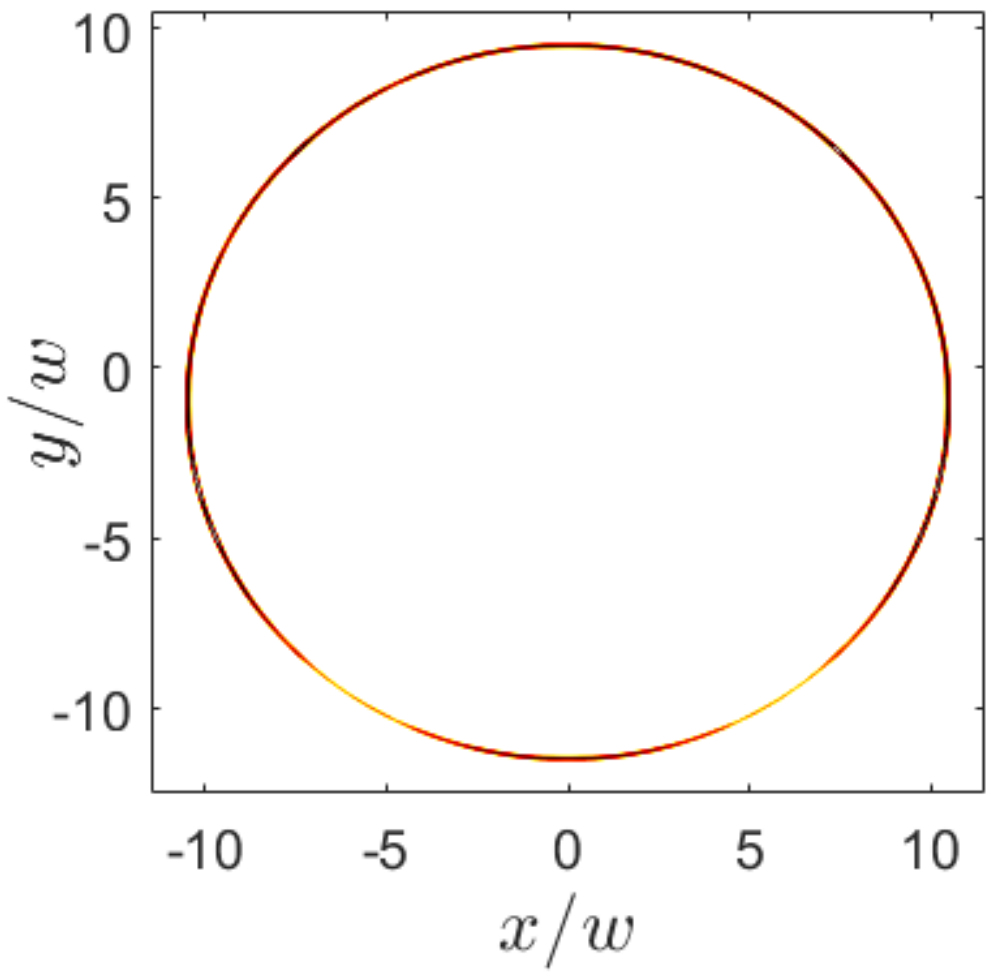}
\includegraphics[width=0.25\textwidth]{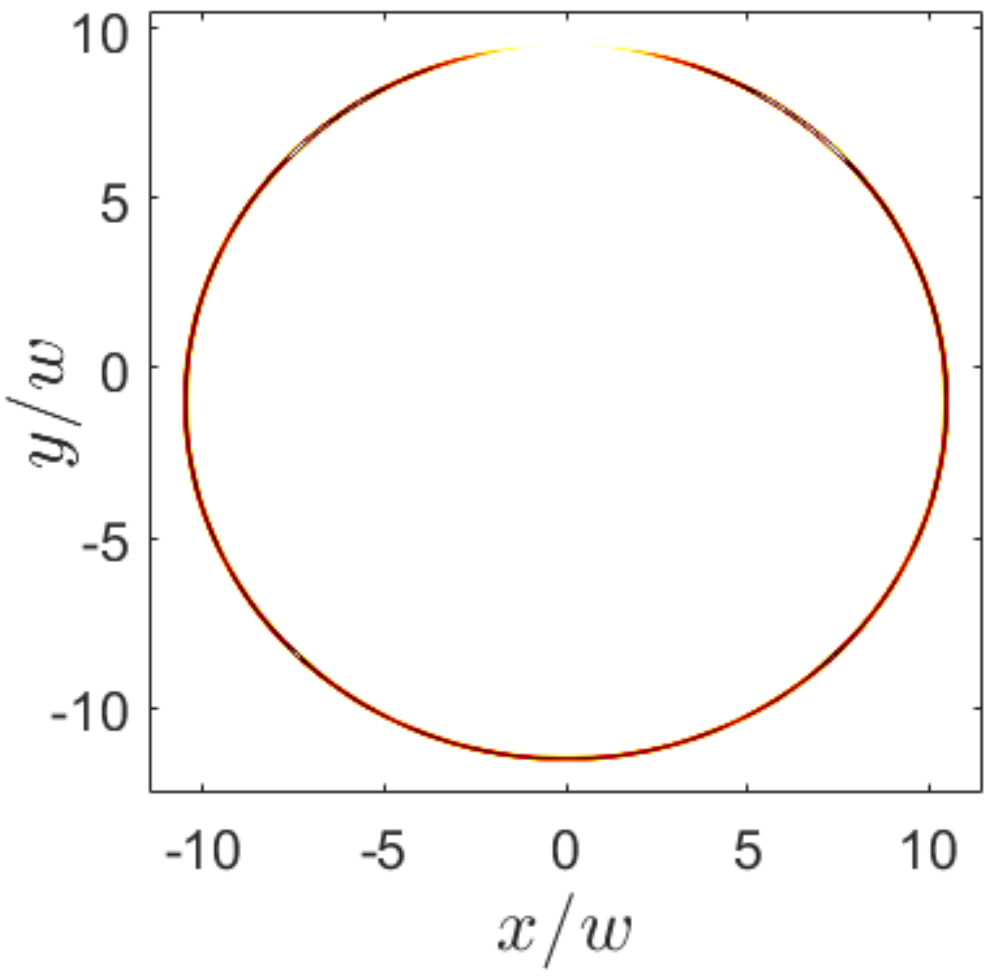}\\
\includegraphics[width=0.05\textwidth]{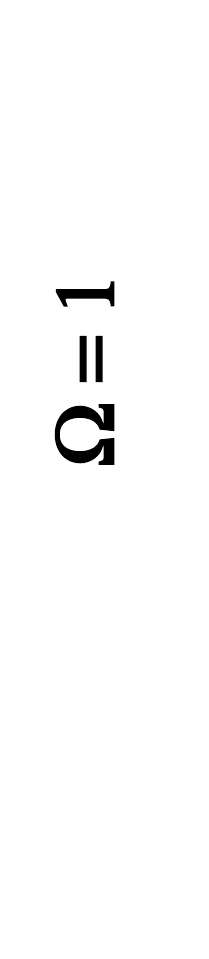}
\includegraphics[width=0.25\textwidth]{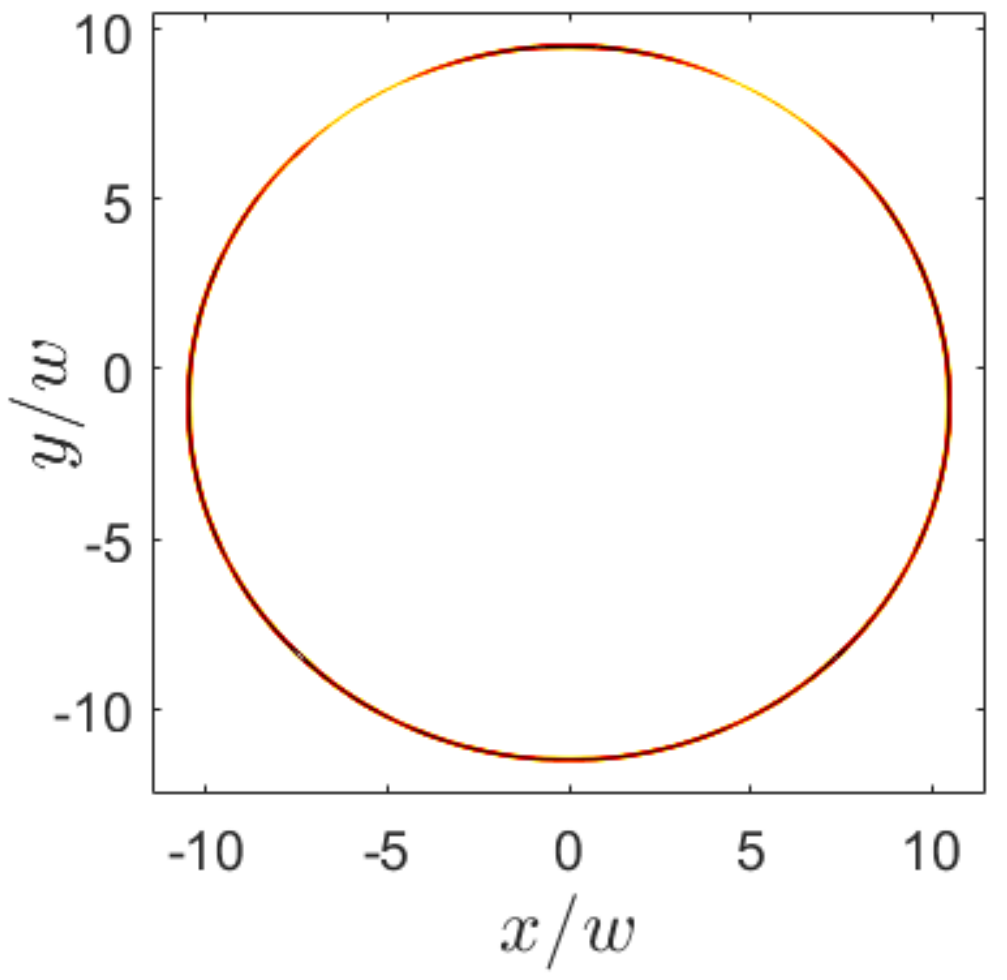}
\includegraphics[width=0.25\textwidth]{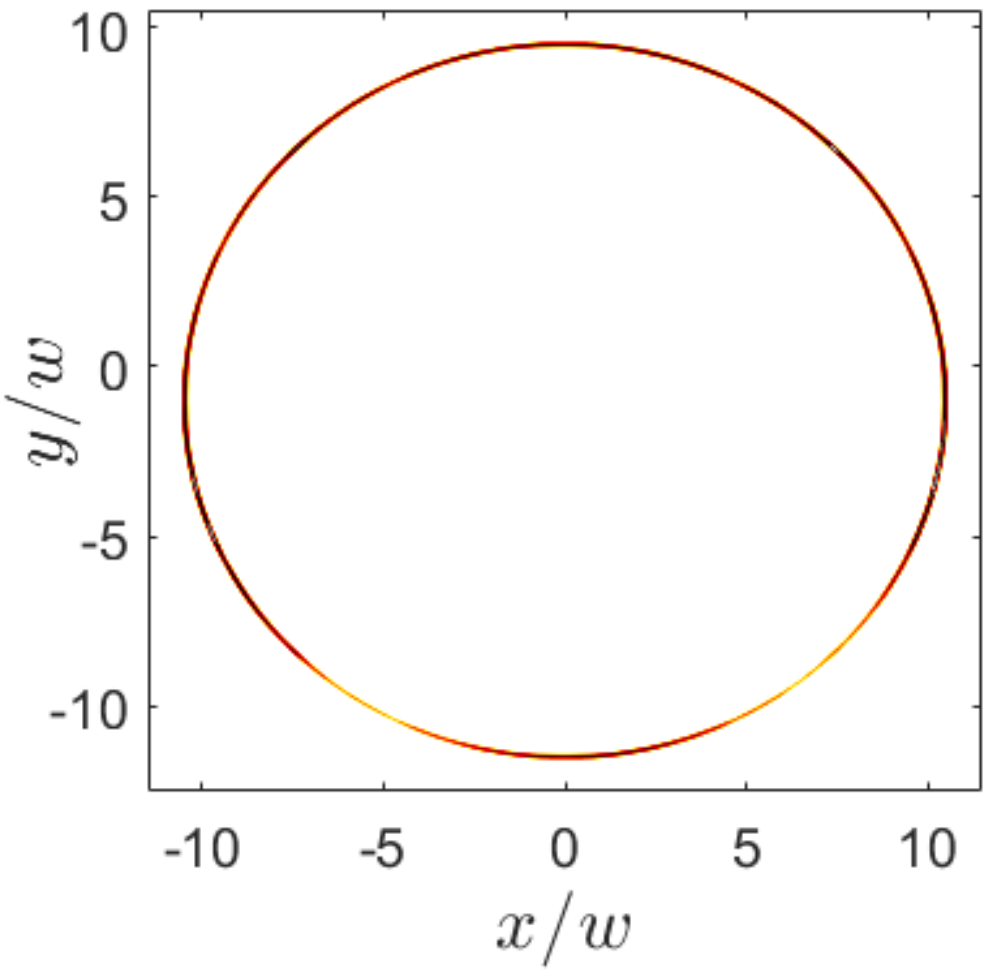}
\includegraphics[width=0.25\textwidth]{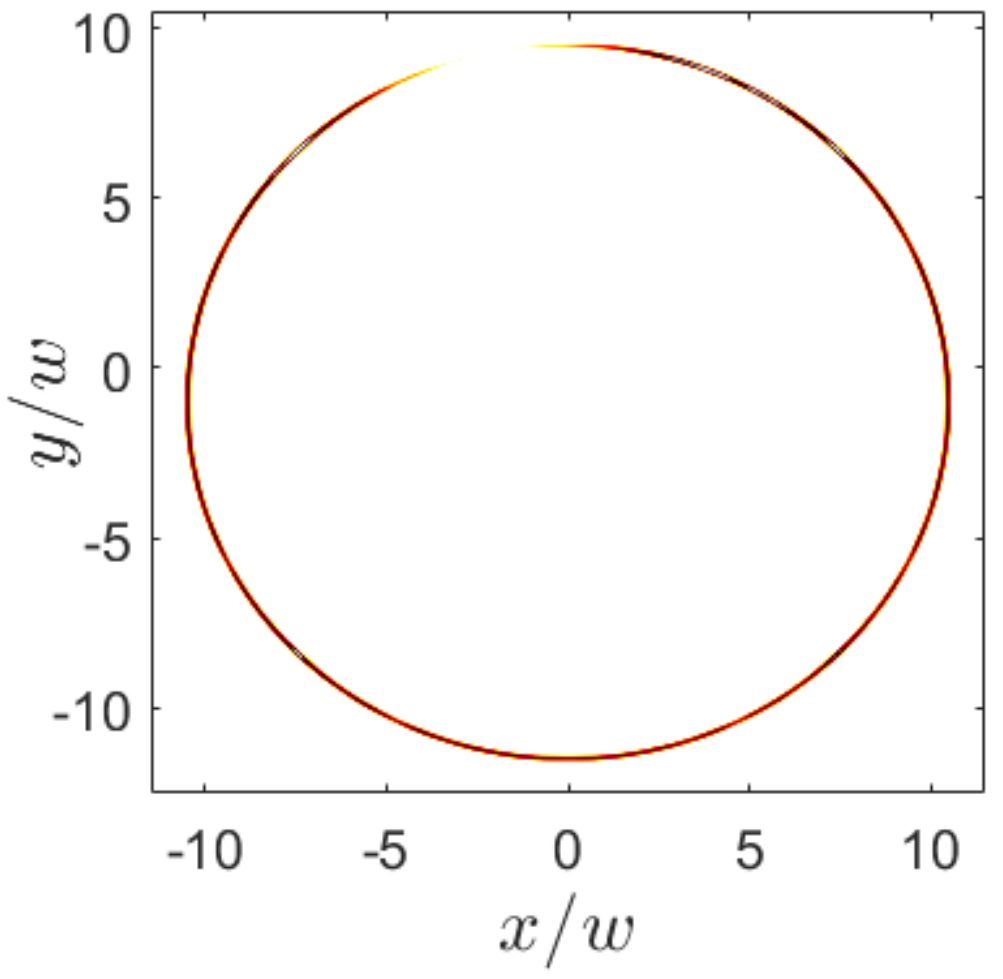}
\caption{Dynamics of the perturbative density modulations (lighter regions on the ring) for the non-rotating 
case (top panels) and rotating case with $\Omega=+1$ (bottom panels). From left to right $t=16, 48$ and $110$. 
In the non-rotating case the modulations move with same speed in the ring and meet at the tunneling point where 
they were produced. For the rotating case, the modulation which moves counterclockwise is faster. It passes 
tunneling point and reaches the slower modulation on the left side of the tunneling point before the slow one 
can reach the tunneling region. Here only the tip of the atomic density with value between $0.0212$ to $0.022$ 
($0.022$ being the maximum of density) are plotted. The white area has density lower than $0.0212$. Perturbations 
are around $3\%-4\%$ of the maximum density.
\label{fig:doppler}}
\end{figure*}
\end{widetext}

As an example of the evolution of density perturbations in the system studied in main text, Fig.~\ref{fig:doppler} shows 
the location of the density modulations for the cases with $\Omega=0$ (top panels) and $\Omega=+1$ (bottom panels) at times 
$t=16, 48$ and $t=110$ when the two counter-rotating density modulations have met. The meeting point for the rotating case 
is clearly dragged along the supercurrent in the ring. For $\Omega=\pm 1$, the two meeting points are symmetrically tilted 
with respect to the one for $\Omega=0$. 

The exact value of $v_0$ and therefore $T_0$ and $T^{\pm}$ depend on the details of the excitations and the sound velocity in the 
system and we are not able to calculate them exactly for our system. However, having an estimation of the $T_0$ makes it possible 
to calculate $T^{\pm}$ and have an estimation of time delay between the fast and slow moving perturbations. In Fig.~2 of the main text, 
for the case with $\Omega=0$, the time when the first large dip in $N_{tot}$ takes place is an approximate value of $T_0$. In our 
system this time is $T_0=116$. Table~\ref{tab:T} summarizes the analytical prediction of $T^{\pm}$, based on Bogoliubov calculations 
and numerical estimation of $T_0$, as well as numerical values extracted from Fig.~2 of the main text.

\begin{table}
  \begin{center}
    \begin{tabular}{|c||c|c||c|c|}
     \hline
		   & \multicolumn{2}{|c||}{\text{Analytical}} & \multicolumn{2}{|c|}{\text{Numerical}}\\
			\hline
       $\Omega$ & \textbf{$T^+$} & \textbf{$T^-$} & \textbf{$T^+$} & \textbf{$T^-$}\\
      \hline
      $\pm 1$ & $112.193$ & $120.075$ & --------- & ---------\\
			\hline
      $\pm 2$ & $108.628$ & $124.446$ & $110$ & $125$\\
			\hline
      $\pm 3$ & $105.282$ & $129.147$ & $108$ & $129$\\
     \hline
    \end{tabular}
		\caption{The analytical estimation of $T^{\pm}$, based on Bogoliubov calculations and numerical estimation of $T_0$,
		compared with the values estimated from numerical results plotted in Fig.~2 of the main text.
    \label{tab:T}}
  \end{center}
\end{table}
 
In conclusion, the one-dimensional calculations based on Bogoliubov excitations together with our rough estimation of the value of 
$T_0$, predict a time delay of $15.818$ and $23.865$, for cases with $\Omega=\pm 2$ and $\Omega=\pm 3$ respectively, between the 
first arrival of the slow and fast density perturbations at the tunneling point. Our numerical data show delays of $15$ and $21$ 
respectively. The predicted values of $T^{\pm}$ for the case with $\Omega=\pm 1$ have not been resolved in our numerical data, due 
to the finite length of the density perturbations and limited time resolution of our saved data. However, the minimum in $N_{tot}$ 
for this case takes place around $t=115$ which is still earlier than $T_0$ and moreover the dip is much shallower and slightly wider 
than the non-rotating case. We attribute the discrepancy between the analytical and the (estimated) numerical $T_0$ to the finite residing 
time (time in which the suppression of the density stays localized, before the pair excitations start) that we observe to characterize 
the decay of the excitations.
\end{appendix}

%

\end{document}